\documentclass[a4paper,12pt]{article}

\usepackage{amsmath}    
\usepackage{amssymb}    
\usepackage{bbm}        
\usepackage{graphicx}   
\usepackage{verbatim}   
\usepackage{color}      
\usepackage[margin=1in]{geometry}
\usepackage{fancyvrb}   
\usepackage[usenames,dvipsnames]{xcolor}   
\usepackage[latin1]{inputenc}  
\usepackage{natbib}     
\usepackage{threeparttable}
\usepackage{url}
\usepackage{relsize}    

\usepackage{tikz}       
\usetikzlibrary{arrows,automata, decorations.text}
\usepackage{pgf}
\usepackage{mathtools} 

\newcommand {\be} {\begin {equation} }
\newcommand {\ee} {\end {equation} }
\newcommand {\bea} {\begin {eqnarray} }
\newcommand {\eea} {\end {eqnarray} }
\newcommand {\dn}[1] {\boldsymbol #1}

\newcommand {\ldt} {{\lambda_T}}
\newcommand {\ltil} {\tilde{\lambda}_T}
\newcommand {\ystar} {{\mathbf{y}^*}}
\newcommand {\nstar} {{n^*}}

\newcommand{\Keywords}[1]{\par\noindent{{\em Keywords\/}: #1}}

\usepackage{multirow}
\usepackage{booktabs}
\newcommand{\ra}[1]{\renewcommand{\arraystretch}{#1}}

\DeclareFontFamily{OT1}{pzc}{}
\DeclareFontShape{OT1}{pzc}{m}{it}{<-> s * [1.10] pzcmi7t}{}
\DeclareMathAlphabet{\mathpzc}{OT1}{pzc}{m}{it}

\begin{document}

\author{
E. ~Charitidou\thanks{E.~Charitidou is with the Department of
Mathematics, National Technical University of Athens, Zografou
Campus, Athens 15780 Greece; email
\texttt{echarit@central.ntua.gr}}, \
D.~Fouskakis\thanks{D.~Fouskakis is with the Department of
Mathematics, National Technical University of Athens, Zografou
Campus, Athens 15780 Greece; email
\texttt{fouskakis@math.ntua.gr}} \
and I.~Ntzoufras\thanks{I.~Ntzoufras is with the Department of
Statistics, Athens University of Economics and Business, 76
Patision Street, Athens 10434 Greece; email
\texttt{ntzoufras@aueb.gr}} \
}
\title{\textbf{Bayesian transformation family selection: moving towards a transformed Gaussian universe}}
\date{}
\maketitle



\setlength\parindent{0pt}      
\setlength{\parskip}{0.15in}   







\section*{Summary}

The problem of transformation selection is thoroughly treated from a Bayesian perspective. 
Several families of transformations are considered with a view to
achieving normality: the \textit{Box-Cox}, the \textit{Modulus}, the \textit{Yeo \&
Johnson} and the \textit{Dual} transformation. 
Markov chain Monte Carlo algorithms
have been constructed in order to sample from the posterior distribution of the
transformation parameter $\lambda_T$ associated with each competing family $T$. We
investigate different approaches to constructing compatible prior distributions for
$\lambda_T$ over alternative transformation families, using a unit-information power-prior
 approach and an alternative normal prior with approximate unit-information interpretation. 
Selection and discrimination between different transformation families is attained via posterior model probabilities. 
We demonstrate the efficiency of our approach using  a variety of simulated datasets. 
Although there is  no  choice of transformation family that can be universally
applied to all problems, 
empirical evidence suggests that some particular data structures
are best treated by specific
transformation families. For example, skewness is associated with  the  Box-Cox family while fat-tailed 
distributions are efficiently treated using the Modulus transformation. \vspace{2mm}

\Keywords{Bayesian model selection;  MCMC;  Posterior model probabilities; Power-prior; Prior
compatibility; Transformation family selection; Unit-information prior.}

\section{Introduction}
\label{sec:introduction}

The pursuit of the optimal transformation for a variable  is considered to be of great concern  as it suits a variety of purposes. 
Normality is a  fundamental assumption of a standard linear model when it comes to the model errors,  along with the assumptions of additive error structure and homoscedasticity. In addition, it is a prerequisite of conjugacy in Bayesian analysis leading to simpler computations. Within the Bayesian  framework, we address the problem of transformation family selection.  Several parametric families of transformations are considered aiming towards the normality of a  response $Y$. In particular, the \textit{Box-Cox} \cite[]{box_cox_64}, the \textit{Modulus} \cite[]{john_draper_80}, the  \textit{Yeo \& Johnson} \cite[]{yeo_johnson_2000} and the \textit{Dual} \cite[]{yang_2006} transformation families are considered in this article.

Several researchers have delved into the area of Bayesian transformed modeling.    
\cite{pericchi_81} considered the linear regression model  transformed as in \cite{box_cox_64} and  specified a non-informative  prior  which is not outcome dependent. In this manner, he managed not only to derive the optimal transformation value  associated with normality but also to account for the assumptions of homoscedasticity and additivity. 
\cite{sweeting_84, sweeting_85} also investigated  a non-empirical prior of the basic transformation parameter $\lambda$ under the Box-Cox family when having vague prior information on the rest of the model parameters but  claimed to deal with some unwanted properties in Pericchi's reasoning. He was mainly concerned with the problem of non-identifiability in a neighbourhood of $\lambda$ taking into consideration that the  model parameters should be a priori independent of $\lambda$ at any value $\lambda_0$ in a neighbourhood of $\lambda$.
In their article \cite*{hoeting_etal_2002} dealt with multivariate problems within the linear model framework and proposed simultaneous variable and transformation selection. The explanatory variables were transformed via a change-point transformation whereas the Box-Cox family was employed for the response. Since their main focus was  the optimization of the predictive performance, Bayesian model averaging was  applied through a $MC^3$ algorithm to best treat model uncertainty. 
An interesting approach was proposed by \cite{gottardo_raftery_2009}    combining model selection, transformation selection and outlier identification simultaneously, again under the Box-Cox family. To escape the problematic nature of inference under transformation, generalized regression coefficients were introduced.
These transformation-free parameters  have a similar interpretation to the usual regression coefficients on the original scale of the data.

In the literature, the term \emph{transformation selection} so far pertains to the  choice of an optimal value of the transformation parameter within a particular family (mostly the Box-Cox family).
To our knowledge, there has been no published research, neither Bayesian nor frequentist, evaluating and/or comparing different  transformation families. Our contribution on the subject relies in extending the meaning of transformation selection to incorporate the latter procedure. In particular, we introduce a two-step approach where a transformation family is selected at an initial level and at a second level the value of the transformation parameter is specified given the selected family. Working within the Bayesian context requires careful choice of  prior distributions. In our case, this becomes even more complex since the prior distributions for the transformation parameter $\lambda_T$ under each transformation family $T$ need to be compatible with each other to account for the different interpretation of $\lambda_T$ given $T$.  Hence, prior compatibility is a fundamental issue  in our transformation selection problem.

Section \ref{sec:transformationfamilies} introduces the transformation families of interest for this study and reveals differences and similarities among them. Section \ref{sec:bayesianformulation} unfolds the approach of  Bayesian inference and transformation selection. Prior specification is presented in detail in this section based on  two different approaches. Computational details for the calculation of the marginal likelihood are provided in Section \ref{sec:computational} with special focus on \citeauthor{chib_95}'s \citeyearpar{chib_95} estimation method. Section \ref{sec:illustrations} exhibits  applications of various simulated datasets picked precisely for illustrative purposes.  Section \ref{sec:discussion} contains the final discussion and possible extensions under consideration.


\section{Transformation Families}
\label{sec:transformationfamilies}

Four families of transformations are considered and contrasted with each other: the \textit{Box-Cox} (BC), the \textit{Modulus} (Mod),  the \textit{Yeo \& Johnson} (YJ) and the \textit{Dual}  transformation. 
 All of them are uni-parametric transformations, in the sense  that they contain only one unknown transformation parameter.  
 The corresponding parameter space  can be either discrete or continuous. The former has the advantage of being more interpretable and less complex (in computational terms as well) while the latter usually results in more accurate choices.
Note that the terms \emph{transformation family} and \emph{transformation class} are used interchangeably throughout this article.

Each family is indexed by a transformation indicator $T$ and involves a transformation parameter  $\lambda_T$. Let us denote by $\mathbf{y}=(y_1,\ldots,y_n)^{\mathrm{T}}$ the observed data and by   
$\mathbf{y}^{(\ldt)}=\big(y_1^{(\ldt)},\ldots,y_n^{(\ldt)}\big)^{\mathrm{T}}$  the transformed ones for a given $\lambda_T$ within a particular transformation family $T$. We aim to identify which   $\mathbf{y}^{(\ldt)}$ can be safely assumed to be a sample from  a normal distribution  with parameters $\left(\mu_T,\sigma^2_T\right)$  under  some appropriate value of the transformation parameter $\lambda_T$.

For any given family $T$, the likelihood of the original data $\mathbf{y}$  is fully specified via the  inverse transformation $\mathbf{y}^{(\ldt)}\rightarrow \mathbf{y}$; thus it consists of the  likelihood of the transformed data multiplied by the absolute value of the determinant of the associated  Jacobian matrix  $\big\vert J(\mathbf{y},\lambda_T|T)\big\vert=  \prod_{i=1}^n{\Big\vert\frac{\partial{y_i^{(\lambda_T)}}}{\partial{y_i}}\Big\vert}$. Thus the likelihood is given by

\begin{equation}
f\left(\mathbf{y}\vert \mu_T, \sigma_T^2, \lambda_T,T\right)= \left(2 \pi \sigma_T^2\right)^{-\frac{n}{2}} \exp \left(-\frac{1}{2 \sigma_T^2} \sum_{i=1}^{n} \left(y_i^{(\lambda_T)} -  \mu_T \right)^2 \right) \times \prod_{i=1}^n \Bigg\vert\frac{\partial{y_i^{(\lambda_T)}}}{\partial{y_i}}\Bigg\vert.
\label{eqn:fullloglikelihood}
\end{equation}



The  formulas of the transformation families compared in this article along with the determinant of their respective Jacobian terms $\vert J(\mathbf{y},\lambda_T|T)\vert$ in absolute value are presented in Table \ref{tbl:transformationfamilies}. The \textit{Identity} (Id) and \textit{Logarithmic} (Log) transformations have been also included 
in the set of models under comparison.

The  renowned paper  of \cite{box_cox_64} describes a  simple and easy-to-use parametric class of variable transformations. One of the primary advantages of the Box-Cox (BC) power class of transformations is that the corresponding Jacobian term   is easily calculated and so is the likelihood function in relation to the original observations.  This class is an extension of the much simpler monotonic function of \cite{tukey_57} which nonetheless had a discontinuity at $\lambda_T=0$. 
A constraint of the Box-Cox transformation is that each observation $y_i,\, i=1,\ldots,n$, is assumed to lie in the strictly positive range of values, zero not included. When dealing with data in $\mathbb{R}$, the simplest solution is to shift the data to the right by adding a large enough shifting quantity $\xi> \min (\dn{y})$.   A simple approach would be to arbitrarily set the shifting parameter equal to $\xi=\vert \min(\dn{y})\vert +\epsilon$, where $\epsilon$ represents a small positive quantity. No shifting corresponds to $\xi=0$.

Several arguments have been laid against the shifted Box-Cox approach; one of them is that asymptotic results of the maximum likelihood theory may not be appropriate to use since the range of the transformed variable depends on the shifting constant, the selection of which is somewhat arbitrary \cite[]{yeo_johnson_2000}. Moreover, the choice of the shifting constant is likely to affect the choice of $\lambda_T$.

Overcoming the obligatory positiveness of the observed data, \cite{john_draper_80} introduced the Modulus transformation (Mod)  which is a monotonic transformation family that seems to work when some sort of symmetry already exists. By substituting $y_i$ with $y_i+1$ we pass from the Box-Cox to the Modulus family for positive $y_i$.

\begin{table*}
\caption{The six transformation families and the determinants of their associated Jacobian terms $\big\vert J(\mathbf{y},\lambda_T|T)\big\vert$ in absolute value. Where not specified, $y_i\in\mathbb{R}$.}  \vspace{2mm}
\centering
\small
\ra{1.5}
\tabcolsep=0.06cm
\begin{tabular}{@{}lllcl@{}}\toprule
Family $T$ & \phantom{ab} & $y_i^{(\ldt)}$ & \phantom{ab} &  $\big\vert J(\mathbf{y},\lambda_T|T)\big\vert$ \\ 
\cmidrule{1-1} \cmidrule{3-3} \cmidrule{5-5}
Id && $=y_i$ &&    $=1$ \\
		\multirow{5}{*}{}\\[-2ex]

Log && $=\log(y_i) \,$ , \quad $y_i>0$ &&  $=\prod\limits_{i=1}^n\left(y_i^{-1}\right)$  \\
		\multirow{5}{*}{}\\[-2ex]

  Box-Cox && $=\begin{cases}
 	                               \frac{y_i^\ldt -1}{\ldt}\, , & \quad \ldt\neq 0\\
                                 \log(y_i)\, , & \quad\ldt=0
                                \end{cases} \quad y_i>0$  &&   $=\prod\limits_{i=1}^n\left(y_i^{\lambda-1}\right) $ \\
		\multirow{5}{*}{}\\[-2ex]

  Modulus &&  $=\begin{cases}
 \frac{sign(y_i)\big[ \left( \vert y_i\vert +1\right)^\ldt-1\big]} {\ldt} , &  \ldt\neq0 \\
 sign(y_i)\log\left( \vert y_i\vert +1\right) , & \ldt=0
\end{cases}$ && $=\prod\limits_{i=1}^n\big(\vert y_i\vert +1\big)^{\ldt-1}$  \\
		\multirow{5}{*}{}\\[-2ex]

  Yeo \& Johnson &&  $=\begin{cases}
 \frac{\left(y_i+1\right)^\ldt-1}{\ldt} \, ,   &    y_i\geq0,   \ldt\neq0 \\
 \log(y_i+1) \, ,    &    y_i\geq0,   \ldt=0     \\
 -\frac{\left(-y_i+1\right)^{2-\ldt}-1}{2-\ldt}  \, ,  &    y_i<0,   \ldt\neq2 \\
 -\log(-y_i+1) \, , & y_i<0, \ldt=2 \\
\end{cases}$ && $=\begin{cases}
 \prod\limits_{i=1}^n\left(y_i +1\right)^{\ldt-1} \, ,  & \quad   y_i\geq0 \\
 \prod\limits_{i=1}^n\left(-y_i+1\right)^{1-\ldt}  \, ,  & \quad   y_i<0  \\
              \end{cases}$ 	\\  
    \multirow{5}{*}{}\\[-2ex]

	Dual &&   $=\begin{cases}
 \frac{y_i^{\ldt}-y_i^{-\ldt}}{2\ldt} \, , & \ldt>0\\
 \log(y_i) \, , & \ldt=0
\end{cases}\quad y_i>0$ &&  $ =\prod\limits_{i=1}^n\frac{y_i^{\ldt-1}+y_i^{-\ldt-1}}{2} $   \\ 
\bottomrule
\end{tabular}
\label{tbl:transformationfamilies}
\end{table*}


\begin{figure}[tb]
\centering
\begin{tikzpicture}[->,>=stealth',shorten >=1pt,auto,node distance=3.8cm,
                    semithick]
  \tikzstyle{every state}=[fill=gray,draw=none,text=white]
  \tikzset{label1/.style={sloped,above} ,
           label2/.style={sloped,below} }

  \node[state]           (A)                    {Modulus};
  \node[state,fill=none] (B) [above right of=A] {};  
  \node[state]           (D) [below right of=A] {Id};
  \node[state]           (C) [below right of=B] {Yeo-Johnson};
  \node[state]           (E) [below of=D]       {Box-Cox};
  \node[state]           (F) [below right of=E] {Dual};
  \node[state]           (G) [below left of=E]  {Log};
  
  \path (A) edge              node[text width=6cm] {\qquad\qquad \qquad\qquad $y>0$}    (C)
            edge[bend left]   node[label1] {\quad $y<0$,\, $\lambda\coloneqq\lambda-2$}         (C)
            edge              node[label1] {$\lambda\coloneqq 1$}                                (D)        
        (C) edge              node {}                                                   (A)
            edge              node[label1] {$\lambda\coloneqq 1$}                                (D)
            edge [bend left]  node[label2,yshift=0em] {$y>0$,\, $y\coloneqq y-1$}                (E)
        (E) edge [bend left]  node[label2] {$y>0$,\, $y\coloneqq y+1$}                           (A)   
            edge              node[label1] {$\lambda\coloneqq 0$}                                (G)  
        (F) edge              node[label1] {$y^{-\lambda}\leftrightarrow 1$}                 (E)
            edge              node[label2] {$y^{(\lambda)}\leftrightarrow 2y^{(\lambda)}$}   (E)
;
\end{tikzpicture}
\caption{Relationships between the transformation families under study.}
\label{fig:transformations}
\end{figure}
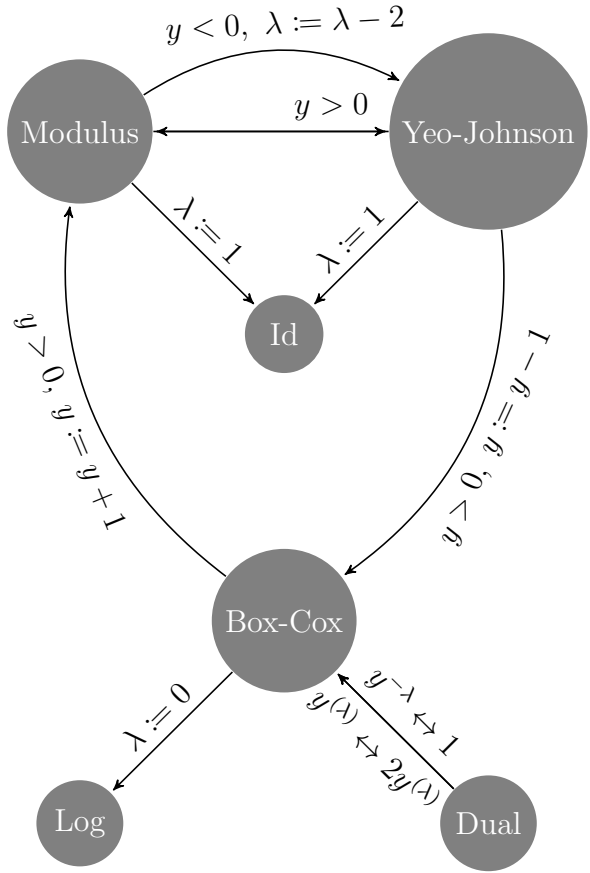

Another modification of the standard Box-Cox power transformation was suggested in the article of \cite{yeo_johnson_2000} with particular interest in mitigating the skewness of a distribution based on the notion of relative skewness \cite[p.3]{vanzwet_64}. This modification also attempts to correct for the problem of restrictive values of the original data. 
The form of the Yeo \& Johnson (YJ) transformation is a smooth alternative to the Modulus transformation.  For positive $y_i$'s, the transformation is identical to the Modulus transformation, thus also equal to Box-Cox if  each $y_i$ is substituted with $y_i-1$. 

Another very interesting idea was eloquently described by \cite{yang_2006} where, once again, only positive observations are considered. The Dual transformation is said to overcome the problem of truncation of the transformed data  by extending the bound; therefore there is no neutral value of $\lambda_T$ in contrast to the common value of one that corresponds to no power transformation at all for the rest of the transformation families  examined in this work. Empirical evidence,
based on normally distributed datasets of various sizes, has estimated 
 $\lambda_T$ to lie approximately  on the interval $1.05-1.30$.
For  values of $\lambda_T$ close to zero, the Dual  approaches the Box-Cox transformation. 
Due to symmetry of the transformation function around $\lambda_T=0$, only positive values of the parameter are considered. 

Figure \ref{fig:transformations} illustrates the relationships between the various transformations in a more elegant and compact way. Note that the shifting procedure may also take place for the Dual and the Log transformations if necessary.



\section{Bayesian Formulation}
\label{sec:bayesianformulation}

In this section, we discuss the Bayesian formulation of the transformation selection problem. 
Focus is given on the Bayesian inference for particular transformations and for particular families and also on the prior specification and  on the derivation of the posterior distribution. 

Concerning model selection, a two stage process is followed. Primarily, we choose the  best transformation family for a given dataset and at a second level, we select the optimal value of the transformation parameter $\ldt$ given $T$.

\subsection{Bayesian Inference for Specific Transformations}
\label{sec:bayesianformulation1}

Focus is given on the inference of the appropriate transformation to achieve normality, 
thus $(\mu_T, \sigma^2_T)$  are regarded as nuisance parameters. 
 Following a similar approach as in 
\cite{berger_pericchi_varshavsky_98} and 
\cite{berger_bernardo_sun_2009} 
we use the (improper) independence Jeffreys (reference) prior for $(\mu_T, \sigma^2_T)$,  
 i.e. 
\begin{equation}
\pi^N\left(\mu_T,\sigma^2_T\vert T\right) \propto \frac{1}{\sigma^2_T}~. 
\label{eqn:musigmajeffreys}
\end{equation}
The Bayesian comparison between different models through the use of posterior model probabilities with such an improper
prior is justified since $(\mu_T, \sigma^2_T)$  are location and scale parameters that appear in every model under comparison \citep{maruyama_strawderman_2013}
and the unknown  normalizing constant in (\ref{eqn:musigmajeffreys}) is common for all $T\in \cal{T}$.

Under any transformation $T$ with specific parameter $\lambda_T$, the  likelihood of the original data marginalized on $\ldt$ is equal to:
\begin{eqnarray}
f(\mathbf{y}\vert\lambda_T,T) 
&=& \int f\left(\mathbf{y}\big\vert \mu_T, \sigma_T^2, \lambda_T, T \right) f( \mu_T, \sigma_T^2 \big\vert  T )  d\mu_T d\sigma_T^2 \nonumber \\
&=& f\left(\mathbf{y}^{(\lambda_T)}\big\vert T\right) \times \prod_{i=1}^n \Bigg\vert\frac{\partial{y_i^{(\lambda_T)}}}{\partial{y_i}}\Bigg\vert \nonumber \\
&\propto& {\big(S^2_T\big)}^{-\frac{n-1}{2}} \times \prod_{i=1}^n \Bigg\vert\frac{\partial{y_i^{(\lambda_T)}}}{\partial{y_i}}\Bigg\vert
\label{eqn:loglikelihood}
\end{eqnarray}
where $S^2_T$ is the sample variance of the transformed data.


\subsection{Bayesian Inference for Parametric Transformation Families}
\label{sec:priorspecification}

Within the Bayesian context, the construction of a prior distribution,  for both the model space indicator  
 $T\in\mathcal{T}$ and the parameter  vector $\boldsymbol{\theta}_T=\left(\mu_T,\sigma^2_T,\lambda_T\right)^\mathrm{T}$, 
 is of paramount importance for the modeling process.

Regarding the prior probability of each of the six  transformations, we use a discrete uniform distribution on 
$\cal{T}$ to express our prior ignorance:
\begin{equation}
\pi(T)= \frac{1}{|\cal{T}|}=\frac{1}{6} \, .
\label{eqn:Tprior}
\end{equation}
\noindent 
For the prior on the transformation parameters, we use the following  structure  
 $$
 \pi(\boldsymbol\theta_T\vert T) = \pi^N\left(\mu_T,\sigma^2_T\big\vert T\right)\pi(\lambda_T \vert T).  
 $$ 
 where $\pi^N\left(\mu_T,\sigma^2_T\big\vert T\right) \propto \sigma_T^{-2}$ as explained in Section \ref{sec:bayesianformulation1}.

Concerning $\ldt$, we propose the use of two distinct prior distributions. 
Two key issues are taken into consideration to form these priors. 
Firstly,
there are   compatibility issues concerning the selection of a prior for $\ldt$ due to the different  interpretation of this parameter  among the transformation families. For instance, the meaning of $\ldt=0$ varies among families and may correspond to the logarithm of the original data or to the negative logarithm of shifted data; see Table~\ref{tbl:transformationfamilies}. Similarly, the value of $\ldt=1$ may correspond to the Identity transformation  for the Modulus and the YJ families or simply to a shifting of the  data by some quantity or even to a specific transformation of the data (different than the Identity) for the Dual family. Therefore, priors for $\ldt$ should share a common basis for all $T \in \cal{T}$. 
The well-known  Lindley-Bartlett paradox \cite[]{lindley_57, bartlett_57} is another aspect that requires caution as to prior selection. 
Model comparison is sensitive to the choice of the prior variance, since very large dispersion  is likely to beget misleading results, 
fully supporting the parsimony principle, that is, the Identity and/or Logarithmic transformations 
regardless what the data suggest for this particular problem. 

On the above  grounds, the concept of the power-prior is adopted by  introducing a set of imaginary data $\mathbf{y}^*$ \cite[]{ibrahim_chen_2000}.
The compatibility between the different transformation families is automatically introduced by the use of common imaginary data since the power-priors are nothing more than rescaled posterior distributions under the assumption of the imaginary data $\ystar$.
Similar strategies have been introduced in common model selection problems, such as the well-known g-prior of \cite{zellner_86}, and in graphical models 
\cite[for example]{ntzoufras_tarantola_2013}.
  Some interesting  properties of this class of  power-priors are described in \cite*{ibrahim_etal_2003}. In addition, a unit-information normal prior (or log-normal prior for the Dual case)   is used in parallel with the previous prior setting, again making use of $\ystar$. This latter approach simplifies computation in half since we evaluate only one integral instead of two, as we show in the sections that follow.


\subsubsection{Power-prior}
\label{sec:powerprior}

In this section, we use the power-prior approach of \cite{ibrahim_chen_2000} to specify our prior for the transformation parameter $\lambda_T$. 
Specifically here we raise $f\left(\ystar \big\vert \ldt, T \right)$,
which is the  likelihood function marginalized on $\ldt$ for  some imaginary  data $\mathbf{y}^*$, 
to a power parameter  $0 < \alpha_0 \le 1$; we call this power-likelihood. 
This  parameter acts as a prior-effect discount parameter. 
We specify $\alpha_0$ to be equal to the inverse of the sample size $n^*$  of the imaginary data so as to enforce a unit-information influence of $\mathbf{y}^*$ on the posterior. 
The power-prior is denoted as prior A.

In order to fully specify the power-prior, we start with a baseline non-informative prior for $\lambda_T$, namely $\pi^N(\lambda_T|T)\propto 1$.
The power-prior of $\lambda_T$ is then proportional to the product of the power-likelihood times the baseline prior.
With the specific choice of  $\pi^N(\lambda_T|T)$ this results to:
\be
\pi_A\left(\lambda_T|\mathbf{y}^*,T\right)= \frac{f\left(\mathbf{y}^*|\lambda_T,T \right)^{1/n^*}}{\mathlarger\int{f\left(\mathbf{y}^*|\lambda_T,T \right)^{1/n^*}\mathrm{d} \lambda_T}}.
\label{eqn:powerprior}
\ee

It often occurs that $\pi_A\left(\lambda_T|\mathbf{y}^*,T\right)$ has no closed form expression and thus the integral involved in the denominator needs to be estimated. 
Extended details on the computation of this quantity are provided in Section \ref{sec:computational}.
As to its shape, the power-prior density lacks symmetry; nevertheless the corresponding  mode is fairly stable and accurate.

Regarding $\ystar$, it ideally  represents available historical data or expert data. In case  neither of those are available, we may consider imaginary data supporting the null hypothesis or some reference model. Here  we propose to use the normal distribution $N\left( \mu_s, \sigma_s^2\right)$ which reflects the Identity transformation that we would ideally like to observe. 
Another approach could be based on using the actual data $\mathbf{y}$ as imaginary resulting to a minimally empirical prior (see \citealp{ntzoufras_2009}). 
In all cases we standardize the original data $\dn{y}$ before being transformed. 
By this way, it is sensible to choose $ \mu_s=0, \sigma_s^2=1$ for the imaginary data. 


\subsubsection{A normal prior with unit-information interpretation}
\label{sec:observedfisher}

With a view of obtaining a closed form expression for the prior in  our case, as opposed to (\ref{eqn:powerprior}),
we  introduce an alternative prior setting (prior B).  
Hence, to simplify the model formulation, we consider a normal prior with (approximate) unit-information interpretation 
as a low information prior. 
By this way, computations of the marginal likelihood become more straightforward since only one integral must be evaluated 
(see Section~\ref{sec:transformationselection} for details).
Hence, under the Box-Cox, the Modulus and the Yeo \& Johnson families, 
we introduce a normal distribution $\pi_B\left(\lambda_T|\ystar,T\right)=N\left(\ldt|{\mu}_{\ldt},{\sigma}_{\ldt}^2,T\right)$  with mean ${\mu}_{\ldt}$ and variance ${\sigma}^2_{\ldt}$. 
As to the Dual family, the normal prior pertains to $\log\ldt$ instead of   $\ldt$ so that the parameter 
under estimation lies in the whole real line. 
Therefore, $\ldt$ has a log-normal prior $LN\left( \ldt| {\mu}_{\log{\ldt}},{\sigma}_{\log{\ldt}}^2,T \right)$.
The prior mean value of one  corresponds to the null hypothesis of normality of  $\mathbf{y}$ at least for the  former three families 
(Box-Cox, Modulus, Yeo \& Johnson). 
 Empirical evidence based on  simulated normal datasets of various sizes suggested
that the  $\ldt$ value corresponding to normality under  the Dual transformation depends much on the shifting constant. 
For the particular examples in Section \ref{sec:illustrations}, the  $\ldt$ value corresponding to normality 
was found approximately equal to $1.2$.
On unification grounds, we introduce a new parameter to be used in the remaining of the current section:
\be
\widetilde{\lambda}_T=\begin{cases}
                  \ldt,  & \text{ for } T= \text{BC, Mod, YJ}  \\
									\log{\ldt}  ,  & \text{ for } T= \text{Dual.}
                  \end{cases}
\ee

Concerning the standard deviation ${\sigma}_{\widetilde{\lambda}_T}$ under $T$, it is based on the observed Fisher information of the parameter of interest for a set of imaginary data $\ystar$ evaluated at the mean of the corresponding transformation parameter, i.e.:
\begin{eqnarray}
{\sigma}_{\widetilde{\lambda}_T}&=&
\left[- \frac{\partial^2 }{\partial \widetilde{\lambda}_T^2} \log f\big(\mathbf{y}^* \big|\widetilde{\lambda}_T,T\big)^{1/n^*} \Bigg| _{\widetilde{\lambda}_T=\widehat{\lambda}_T}\right]^{-\frac{1}{2}},
\label{eqn:observedfisher}
\end{eqnarray}
where $\widehat{\lambda}_T=\log\widehat{\lambda}_D$ under the Dual family and $\widehat{\lambda}_T=1$ in all other cases.
In Equation \ref{eqn:observedfisher},  the likelihood marginalized on $\ldt$  for the imaginary data $\mathbf{y}^*$   is raised to the power of $\left(n^*\right)^{-1}$ 
so as to form a unit-information prior.
Then, the standard deviation   for each family $T$ can be summarized by 
\be
{\sigma}_{\widetilde{\lambda}_T}=\left(-\frac{q_T}{\nstar}+\frac{\nstar-1}{\nstar}\left[ \frac{S^2_{\dn{w}_T-\dn{d}_T}+S_{\dn{z}_T\dn{r}_T}}{S^2_{\dn{z}_T}}-2\left(\frac{S_{\dn{z}_T\dn{w}_T}-S_{\dn{z}_T\dn{d}_T} }{S^2_{\dn{z}_T}} \right)^2 \right]\right)^{-\frac{1}{2}}
\label{eqn:fisherall}
\ee
where		
the sample (unbiased) variance of  $\mathbf{x}$ is denoted by $S^2_{\mathbf{x}}$ and 
the sample covariance between  $\mathbf{x}$ and $\mathbf{y}$ is denoted by $S_{\mathbf{x}\mathbf{y}}$; 
see Appendices~\ref{calculationofsigma} and \ref{calculationofsigmadual}  for the detailed derivation of (\ref{eqn:fisherall}) under Box-Cox and Dual 
(for the rest of the families the derivation is similar as in the Box-Cox transformation and therefore is omitted).
The transformed vector $\dn{z}_T$ is given by  
\be\dn{z}_T=\begin{cases}
					 (\ystar+\xi\mathbbm{1}_\nstar)^{(\ldt=1)},  & \text{ for } T= \text{BC}\\
					 (\ystar+\xi\mathbbm{1}_\nstar)^{(\ldt=\widehat{\lambda}_D)},  & \text{ for } T= \text{Dual}\\
           {\ystar}^{(\ldt=1)},  &\,\text{otherwise}
					\end{cases}; 
\ee
where $\mathbbm{1}_n$ is a vector of length $n$ with all elements equal to one and $\xi$ is the shifting parameter. 
Moreover, we define 
\be
\dn{d}_T=\begin{cases}
					\vert\dn{z}_T \vert, & \text{ for } T= \text{YJ}\\
					          \dn{z}_T,& \text{otherwise}
					\end{cases},
\ee
\be 
q_T =
\widehat{\lambda}_D\sum\limits_{i=1}^{\nstar}{ \frac{(y_i^*+\xi)^{2\widehat{\lambda}_D-2}-(y^*_i+\xi)^{-2\widehat{\lambda}_D-2}+4\widehat{\lambda}_D(y^*_i+\xi)^{-2}\log(y^*_i+\xi)}{\left[\log (y_i^*+\xi\mathbbm{1}_\nstar) \right]^{-1}\left[(y^*_i+\xi)^{\widehat{\lambda}_D-1}+(y^*_i+\xi)^{-\widehat{\lambda}_D-1}\right]^2}}, \, 
\ee
for the Dual transformation or $q_T=0$ zero for the rest of the transformations,  
and 
\be\dn{w}_T=\begin{cases}
					(\ystar+\xi\mathbbm{1}_\nstar)\circ\log(\ystar+\xi\mathbbm{1}_\nstar),\quad & \text{ for } T= \text{BC}\\
					sign({\ystar})\circ(\vert\ystar\vert+\mathbbm{1}_\nstar)\circ\log(\vert\ystar\vert+\mathbbm{1}_\nstar),\quad & \text{ for } T= \text{Mod} \\
					\left(\vert\ystar\vert +\mathbbm{1}_\nstar\right)\circ\log(\vert\ystar\vert +\mathbbm{1}_\nstar), \quad & \text{ for } T= \text{YJ}\\
					\frac{1}{2}\Big[({\ystar+\xi\mathbbm{1}_\nstar})^{\widehat{\lambda}_D}+{(\ystar+\xi\mathbbm{1}_\nstar)}^{-\widehat{\lambda}_D}\Big]\circ\log(\ystar+\xi\mathbbm{1}_\nstar), \quad & \text{ for } T= \text{Dual}
					\end{cases}\ee
									with $sign(\ystar)$ being a vector of elements $\{+1, -1\}$ depending on whether the $i$-th element of $\ystar$ is positive or negative. 
Finally,	$r_T$ is given by 			
\be\dn{r}_T=\begin{cases}
\dn{w}_T\circ\log(\ystar+\xi\mathbbm{1}_\nstar)-2(\dn{w}_T-\dn{z}_T), \quad & \text{ for } T= \text{BC}\\
\dn{w}_T\circ\log(\vert\ystar\vert+\mathbbm{1}_\nstar)-2(\dn{w}_T-\dn{z}_T),  & \text{ for } T= \text{Mod}\\
sign({\ystar})\circ\dn{w}_T\circ\log(\vert\ystar\vert+\mathbbm{1}_\nstar)-2\Big(sign({\ystar})\circ\dn{w}_T-\dn{z}_T\Big),  & \text{ for } T= \text{YJ}\\
\dn{z}_T\circ(\widehat{\lambda}_D)^2\circ\log^2 (\ystar+\xi\mathbbm{1}_\nstar)-(\dn{w}_T-\dn{z}_T), & \text{ for } T= \text{Dual}
\end{cases}
\ee 	
with $\circ$ denoting the Hadamard product for component-wise multiplication of two vectors.


\subsection{Posterior Inference for the Transformation Parameter}
\label{sec:posteriorinference}

The main parameters of inferential interest are $\ldt,\,T$. The  parameters $\mu_T,\,\sigma_T$ are considered as nuisance parameters. 
Given the transformation family $T$, the logarithm of the marginal posterior density of $\ldt$ is given by the following equation:
\begin{equation}
 \log \pi\left(\ldt\vert \mathbf{y},T\right) = \log f\left(\mathbf{y}\vert\ldt,T\right) + \log \pi\left(\ldt\vert \ystar, T\right) + c,  
\end{equation}
where $c$ is the logarithm of the normalizing constant of the posterior distribution of $\lambda_T$.

The first term on the right-hand side of the above equation is the  log-likelihood  of the untransformed data marginalized on the transformation parameter $\ldt$  and is given explicitly in (\ref{eqn:loglikelihood}). 
 The final general form of the log-posterior distribution employed, marginalized on $\ldt$, is the following:
\begin{equation}
  \log \pi\left(\ldt\vert \mathbf{y}, T\right) = \log f\left(\mathbf{y}^{(\ldt)}\big\vert T\right) + {\log \big\vert J\left(\mathbf{y},\ldt|T\right) \big\vert}+\log \pi\left(\ldt\vert \ystar,T\right)+c,\,\, c\in\mathbb{R}.
	\label{eqn:profileposterior}
\end{equation}
The third term on the right of (\ref{eqn:profileposterior}) is the prior distribution of $\ldt$ under family $T$ given  $\ystar$ and varies according to the prior setting used as described in Section \ref{sec:priorspecification}.
In order to simulate from (\ref{eqn:profileposterior}) we have constructed an appropriate random walk Metropolis-Hastings (MH) algorithm.


\subsection{Transformation Selection}
\label{sec:transformationselection}

Within the Bayesian framework, the identification of the best transformation among the six transformation families considered is equivalent (assuming a
zero-one loss function) to finding the transformation $T\in\cal{T}$ with the highest
posterior model probability, defined as
\begin{equation}
\pi\left(T\vert\mathbf{y}\right)=\frac{f\left(\mathbf{y}\vert T\right)\pi(T)}{\sum_{ T \in {\cal T} }{f\left(\mathbf{y}\vert T\right)\pi(T)}}
\end{equation}
where $f\left(\mathbf{y}\vert T\right)$ is the marginal likelihood under transformation $T$ and $\pi(T)$ is the prior distribution of transformation family $T$ given in (\ref{eqn:Tprior}). The marginal likelihood can be further expanded to conveniently include the effect of $\ldt$: 

\begin{equation}
f\left(\mathbf{y}\vert T\right)=\int{f\left(\mathbf{y}\vert\ldt,T\right)\pi\left(\ldt\vert \ystar,T\right)}\mathrm{d}\ldt
\label{eqn:marginallikelihoodgeneral}
\end{equation}
\noindent with $f\left(\mathbf{y}\vert\ldt,T\right)$ being the  likelihood of $\mathbf{Y}$ under family $T$ marginalized on  $\ldt$ 
and $\pi\left(\ldt\vert\ystar, T\right)$ representing the prior distribution of  $\ldt$ given  $T$; see Section~\ref{sec:priorspecification}. 
It is evident that the Id and the Log transformations are not associated with any transformation  parameter $\ldt$ but we have adopted a holistic notation  in the sake of cohesion.  
Hence, $f\left(\mathbf{y}\vert\ldt,T\right)$, under the two latter transformations, is given by (\ref{eqn:loglikelihood}) with $\mathbf{y}^{(\ldt)}$ being the original (yet standardized) data $\mathbf{y}$ or the logarithm of $\mathbf{y}$ respectively.

In the case of the power-prior approach (prior A) for $\ldt$ given $T$, the marginal likelihood  is given via the following formula which involves two  integrals:  
\be
f\left(\mathbf{y}|T\right)=  \frac{ \mathlarger\int{ f\left(\mathbf{y}|\ldt,T\right)f\left(\mathbf{y}^*|\ldt,T \right)^{1/n^*} \mathrm{d} \ldt}}{\mathlarger\int{f\left(\mathbf{y}^*|\ldt,T \right)^{1/n^*}\mathrm{d} \ldt}}.
\label{eqn:powerpriormarginal}
\ee

Additionally, for the alternative unit-information prior approach (prior B) for $\ldt$ given $T$, the corresponding formula of the marginal likelihood is the following:
\be
f(\mathbf{y}|T)=\begin{cases}
 \int{ {f\left(\mathbf{y}|\ldt,T\right)N\left(\ldt|\mu_{\ldt},\sigma^2_{\ldt},T \right)} \mathrm{d} \ldt}, & \text{ for } T= \text{BC, Mod, YJ} \\
\int{ {f\left(\mathbf{y}|\ldt,T\right)LN\left(\ldt|\mu_{\log\ldt},\sigma^2_{\log\ldt},T \right)} \mathrm{d} \ldt}, & \text{ for } T= \text{Dual}
\end{cases}.
\label{eqn:fishermarginal}
\ee

Estimation of the marginal likelihood in  (\ref{eqn:powerpriormarginal}) or (\ref{eqn:fishermarginal})  is achieved through an extension of the  \emph{candidate estimator} of Chib \cite[]{chib_95} as described in \cite{chib_jeliazkov_2001}. Section~\ref{sec:computational} provides all the computational details.


\section{Marginal Likelihood Computation}
\label{sec:computational}

The computation of the intractable integral (\ref{eqn:powerpriormarginal}) or (\ref{eqn:fishermarginal}) is achieved using three distinct estimators. The primary one is the  \emph{candidate estimator} of Chib which is based on the results of a Metropolis-Hastings (MH) algorithm simulating from the posterior distribution of $\ldt$.
Prior to this, we  have also used the Laplace-Metropolis estimator \cite[]{lewis_raftery_97} and  a numerical approximation estimator of the integral in question. 
The use of these alternative procedures was mainly adopted in order to certify the accuracy of the results. Results stemming from all three estimators
seem to converge. The most unstable of the three estimators was found  to be the third one, while the Laplace-Metropolis estimator  deviated from the other two when the  posterior distribution of $\ldt$ was considerably non-symmetric, something which mostly occurs  under the Dual family.

The Laplace-Metropolis (LM) estimator is named after the fact that appropriate MCMC output provides essential quantities which are then inserted into the classic Laplace approximation. The formula of the LM estimator is the following:
\be
\log f(\mathbf{y}|T)\approx \frac{1}{2}\log(2\pi)+\frac{1}{2}\log\left(\sigma^*_{\ldt} \right)^2+\log \pi\left(\lambda_T^*| T\right)+ \log f\left(\mathbf{y}|\lambda_T^*, T\right).
\label{eqn:lmgeneral}
\ee
Additionally,  $\lambda_T^*$ stands for the posterior mode of the  $\left\{\ldt\right\}$ chain, which can be sufficiently approximated by the posterior mean or the median,  and  $\left(\sigma^*_{\ldt} \right)^2$ is the MCMC estimate of the posterior variance  of  $\left\{\ldt\right\}$.

For Chib's estimator, we consider the following basic marginal likelihood identity:
\begin{equation}
\log{f(\mathbf{y}|T)}=\log{f\left(\mathbf{y}\vert\lambda_T^*,T\right)}+\log{\pi(\lambda_T^*|\ystar,T)}-\log{\pi(\lambda_T^*\vert\mathbf{y},T)}
\end{equation}

\noindent where $\lambda_T^*$ is a high-posterior-density value of  $\left\{\ldt\right\}$ and the quantity $\pi\left(\lambda_T^*\vert\mathbf{y},T\right)$  is called the \emph{posterior ordinate}. The posterior ordinate is estimated via the formula:

\begin{equation}
\pi\left(\lambda_T^*\vert\mathbf{y}, T\right)
 = (2\pi k^*)^{-1/2} 
   \frac{ \dfrac{1}{M} \sum\limits_{g=1}^{M} \left[ \min \left\{ 1, \dfrac{ K( \lambda_T^* ) }{ K \big( \lambda_T^{(g)} \big) } \right\}
   \exp\bigg\{-\tfrac{ \left(\lambda_T^*-\lambda_T^{(g)} \right)^2}{2k^*}\bigg\} \right] }
        { \dfrac{1}{J}\sum\limits_{j=1}^{J}{\min \Bigg\{ 1, \dfrac{ K\big( \lambda_T^{(j)} \big) }{ K ( \lambda_T^{*} ) } \Bigg\} } }
\label{eqn:analyticposteriorordinate}
\end{equation}
where 
\begin{eqnarray}
 K( \lambda_T ) =&&
\left[\sum\limits_{i=1}^n{\left(y_i^{\left(\ldt\right)}-\overline{\mathbf{y}^{\left(\ldt\right)}}\right)^2}\right]^{-\frac{n-1}{2}}
\Big\vert J\left(\mathbf{y},\ldt|T\right)\Big\vert  
\nonumber \\ 
&\times&  \left[\sum\limits_{i=1}^\nstar{\left({y^*_i}^{\left(\ldt\right)}-\overline{\mathbf{\ystar}^{\left(\ldt\right)}}\right)^2}\right]^{- \frac{\nstar-1}{2\nstar}} 
\Big\vert J\left(\mathbf{\ystar},\ldt|T\right)\Big\vert^\frac{1}{\nstar}  
\label{eqn:acceptanceprobability}
\end{eqnarray}
for the power-prior setup (prior A). 
For prior B, the second line of (\ref{eqn:acceptanceprobability})  is simply replaced by the kernel of 
the normal prior distribution specified in Section \ref{sec:observedfisher} for all transformation families $T$ 
except for the Dual where the log-normal is used instead. 
Moreover, 
$\lambda_T^{(g)}$ is a random sample of size $M$ from the posterior distribution of $\ldt$ obtained by a random walk MH algorithm, 
while 
$\lambda_T^{(j)}$, $j=1,\ldots,J$, is a random sample  of size $J$ generated from the proposal distribution used in our MH algorithm; 
that is, a sample from a normal distribution $N\left(\ldt\vert\lambda_T^*,k^*, T\right)$ with mean $\lambda_T^*$ 
and variance $k^*$ chosen appropriately to achieve good mixing; see, for example, in \cite{ntzoufras_2009}. 
In the following,  we consider $M$ around $15000-18000$ iterations additional to the  \emph{burn in}
while for $J$ we consider only $2000$ since it refers  to the number of i.i.d. draws from the proposal distribution.


\section{Illustrations}
\label{sec:illustrations}

In order to illustrate  our approach, we use simulated data from a variety of distributions. Results are provided for medium  and large samples sizes, namely $n=100$ and $n=1000$, based on the candidate estimator of Chib for the estimation of the marginal likelihood. 
Note that all data have been standardized prior to transformation. 
Moreover,  all observations have been shifted to the positive axis by adding  the absolute value of the minimum observation plus half the smallest non-zero value $y_0$ of the non-negative data (i.e. $\epsilon=y_0/2$)
for the Box-Cox,  the Dual and the Log transformations.

\subsection{Simulated Examples}

In the first example we simulate data from the standard normal distribution; this example serves as a reference (see Table \ref{tbl:normalpmp}). 
Starting with a  sample size of $n=100$, we observe that under both prior approaches, the Identity transformation is  undoubtedly the winner, as it should be. Specifically, the posterior  probability $P(T=\mathrm{Id}\vert\mathbf{y})$   is  $77\%$ under prior A and $76\%$ under prior B. 
The second model in order of preference is the Box-Cox  model with posterior probability around $9\%$ under both priors and  posterior mode of $\ldt$ around $1.07$, correcting for minor divergence from normality. The YJ and Modulus families follow closely with posterior probabilities around $6\%-8\%$ and posterior mode of $\ldt$ close to unity. 
For the large size dataset ($n=1000$),   the Identity transformation is also indicated as the optimal choice, only now the associated posterior model probabilities have soared to reach the level of  $88\%$ under both prior setups.  Box-Cox  follows with posterior probability around $5\%$ and posterior mode of $\ldt$ about $0.93$ (still very close to unity). The importance of the latter family is almost equal to the Modulus and YJ models in terms of posterior probabilities.
In either case, the Log transformation is indicated as  not suitable since it is less flexible compared to the four parametric transformation families that adapt better to each dataset. Dual also shows to be an outlier for these datasets. 
A strong measure of convergence of results under both prior settings is  the very  small deviation between 
the log-marginal likelihood figures under priors A and B. 
Somewhat larger discrepancies are observed in the case of the Dual family,  
since a log-normal prior is used (instead of normal) under prior setting B.
In general, the optimal $\ldt$ value is very close to one for every parametric family except Dual, confirming that there is little need for an actual transformation. 

\begin{table*}[htb]
\begin{threeparttable}    
\caption{Posterior model probabilities and log-marginal likelihood values  for each trasformation family $T$ along with Monte Carlo estimates for the posterior mode (sd) of $\lambda_T$ for  normal simulated datasets.}  \vspace{2mm}
\centering
\footnotesize
\ra{1.3}
\tabcolsep=0.06cm
\begin{tabular}{@{}llcccccccccc@{}}\toprule
\phantom{a} & N(0,1)  && \phantom{a} &Prior\tnote{1}& \phantom{ab} &Id & Box-Cox & YJ & Modulus & Dual    & Log  \\ 
\cmidrule{2-2} \cmidrule{5-5} \cmidrule{7-12}
  &\multirow{6}{*}{$n=100$} & \multirow{2}{*}{$P(T|\mathbf{y})$} && prior A && 0.77 & 0.09 & 0.07 & 0.06 & $<0.01$ & $<0.01$  \\
  && && prior B && 0.76 & 0.08 & 0.08 & 0.07 &   $<0.01$ & $<0.01$ \\
	\multirow{12}{*}{}\\[-2ex]
	&& \multirow{2}{*}{$\log f(\mathbf{y}|T)$} && prior A && -193.14 & -195.30 & -195.54 & -195.61 & -200.77 & -213.00 \\
  && && prior B && -193.14 & -195.35 & -195.36 & -195.55 & -200.25 & -213.00  \\
		\multirow{12}{*}{}\\[-2ex]
  && \multirow{2}{*}{$\lambda_T$} && prior A && -  & 1.07 (0.20) & 1.07 (0.13) & 1.03 (0.28) & 1.52 (0.21) & -	 \\
  && && prior B && - & 1.07 (0.20) & 1.07 (0.13) & 1.02 (0.28) & 1.50 (0.21) & -\\  \hline
	
				\multirow{12}{*}{}\\[-2ex]
	  & N(0,1) & & \phantom{a} &Prior& \phantom{ab} & Id & Box-Cox & Modulus   & YJ & Log & Dual   \\ 
 \cmidrule{2-2} \cmidrule{5-5}  \cmidrule{7-12}
&\multirow{6}{*}{$n=1000$} & \multirow{2}{*}{$P(T|\mathbf{y})$} && prior A && 0.88 & 0.05 & 0.04 & 0.03 & $<0.01$ & $<0.01$ \\
  && && prior B && 0.88 & 0.04 & 0.04 & 0.03 & $<0.01$ & $<0.01$ \\
			\multirow{12}{*}{}\\[-2ex]
	&& \multirow{2}{*}{$\log f(\mathbf{y}| T)$} && prior A && -3103.69 & -3106.60 & -3106.80 & -3107.25 & -3433.63 & -3439.15 \\
  && && prior B && -3103.69 & -3106.68 & -3106.70 & -3107.04 & -3433.63 & -3437.73 \\
			\multirow{12}{*}{}\\[-2ex]
  && \multirow{2}{*}{$\lambda_T$} && prior A && - & 0.93 (0.06) & 1.08 (0.09) & 0.97 (0.04) & - & 0.01 (0.01)   \\
  && && prior B && - &0.92 (0.07) & 1.08 (0.09) & 0.97 (0.04) & - & 0.01 (0.01)  \\
\bottomrule
\end{tabular}
\label{tbl:normalpmp}
\begin{tablenotes}
    \item[1] prior A: Power-prior (see Section~\ref{sec:powerprior}); prior B: Unit-information normal prior (see Section~\ref{sec:observedfisher}).
  \end{tablenotes}
 \end{threeparttable}
\end{table*}

Next, we present an illustration using simulated samples from a Gamma$(2,3)$ distribution in order  to examine the behavior of our approach on highly skewed data
 (see Table \ref{tbl:gammapmp}). 
The best adapting class for this dataset is clearly the Box-Cox transformation for both the medium and  large sample sizes, with the Identity transformation not supported  as anticipated. Moving from one sample size to the other, the posterior model probabilities for Box-Cox remain over $99\%$ while the corresponding posterior mode of $\ldt$ falls slightly from $0.44$ to $0.35$. Notice how the posterior standard deviation of $\ldt$ is undermultiplied by a factor of $3$ when $n=1000$ compared to $n=100$. For the medium size data, the YJ model with posterior $\ldt$ mode of $0.43$  is attributed a minor weight of $1\%$ which becomes totally negligible for the larger dataset.


\begin{table*}[htb]
\begin{threeparttable} 
\caption{Posterior model probabilities and log-marginal likelihood values for each trasformation family $T$ along with Monte Carlo estimates for the posterior mode (sd) of $\lambda_T$ for  Gamma simulated datasets. }  \vspace{2mm}
\centering
\footnotesize
\ra{1.3}
\tabcolsep=0.06cm
\begin{tabular}{@{}llcccccccccc@{}}\toprule
\phantom{a} & G(2,3)  && \phantom{a} &Prior\tnote{1}& \phantom{ab}  & Box-Cox &  YJ  & Id & Modulus & Log & Dual   \\ 
\cmidrule{2-2} \cmidrule{5-5} \cmidrule{7-12}
  &\multirow{6}{*}{$n=100$} & \multirow{2}{*}{$P(T|\mathbf{y})$} && prior A && 0.99 & 0.01 & $<0.01$ & $<0.01$ & $<0.01$ & $<0.01$ \\
  && && prior B && 0.99 & 0.01 & $<0.01$ & $<0.01$ & $<0.01$ & $<0.01$  \\
	\multirow{12}{*}{}\\[-2ex]
	&& \multirow{2}{*}{$\log f(\mathbf{y}|T)$} && prior A && -182.39 & -188.47 & -193.14 & -195.01 & -195.24 & -198.85 \\
  && && prior B && -182.45 & -188.42 & -193.14 & -194.96 & -195.24 & -197.47 \\
		\multirow{12}{*}{}\\[-2ex]
  && \multirow{2}{*}{$\lambda_T$} && prior A &&  0.44 (0.09)  &  0.43 (0.16) & -& 1.30 (0.26) & - &  0.01 (0.04)  \\
  && && prior B &&  0.44 (0.09)  &  0.43 (0.16) & -& 1.30 (0.25) & - & 0.04 (0.04)  	\\  \hline
	
				\multirow{12}{*}{}\\[-2ex]
	  & G(2,3) & & \phantom{a} &Prior& \phantom{ab}  & Box-Cox & YJ & Log & Dual    & Modulus  &  Id  \\ 
 \cmidrule{2-2} \cmidrule{5-5}  \cmidrule{7-12}
&\multirow{6}{*}{$n=1000$} & \multirow{2}{*}{$P(T|\mathbf{y})$} && prior A && $>0.99$ &  $<0.01$ &  $<0.01$ & $<0.01$ &  $<0.01$ & $<0.01$ \\
  && && prior B && $>0.99$ &  $<0.01$ & $<0.01$ & $<0.01$ & $<0.01$ & $<0.01$ \\
			\multirow{12}{*}{}\\[-2ex]
	&& \multirow{2}{*}{$\log f(\mathbf{y}| T)$} && prior A && -2954.62 & -2993.98 & -3011.49 & -3014.96 & -3102.12 & -3103.69 \\
  && && prior B && -2954.63 & -2993.84 & -3011.49 & -3013.88 & -3102.01 & -3103.69 \\
			\multirow{12}{*}{}\\[-2ex]
  && \multirow{2}{*}{$\lambda_T$} && prior A &&  0.35 (0.03) & 0.31 (0.05)&  - & 0.01 (0.03)  & 0.76 (0.07) & - \\
  && && prior B && 0.35 (0.03)  & 0.31 (0.05)& - & 0.03 (0.03)  & 0.76 (0.07) & -   \\
\bottomrule
\end{tabular}
\label{tbl:gammapmp}
\begin{tablenotes}
    \item[1]  prior A: Power-prior (see Section~\ref{sec:powerprior}); prior B: Unit-information normal prior (see Section~\ref{sec:observedfisher}).
  \end{tablenotes}
 \end{threeparttable}
\end{table*}

Finally, the Student distribution  is used to illustrate the performance of our approach 
for symmetrically distributed data but with fat tails. 
This is of particular interest since the latter characteristic usually induces  failure of  transformation to normality under most families according to  our experience. 
Our example uses a Student distribution  with two degrees of freedom $t_2$ and non-centrality parameter equal to minus one.  
Looking at  Table \ref{tbl:studentpmp}, 
we observe that the supremacy of the Modulus family is unquestionable for this distribution   under both  prior setups. 
Even for the smaller dataset ($n=100$), the posterior  probability of the Modulus transformation  is  $93\%$ assigning a  small weight of around $4\%$ to the Box-Cox family and $1\%$ to  each of the YJ and Id models. For $n=1000$ this figure climbs up to over $99\%$ for Modulus.
The corresponding posterior mode  value of $\ldt$ is about $0.14$ in the former case and $-0.4$ for the large sample size while the corresponding posterior standard deviation is 0.25 and 0.08 respectively.   It is worth mentioning that similar behavior and support of 
Modulus was also observed on simulation studies based on  the Laplace distribution which is another example of a fat-tailed symmetric density.


\begin{table*}
\begin{threeparttable}    
\caption{Posterior model probabilities and log-marginal likelihood values  for each trasformation family $T$ along with Monte Carlo estimates for the posterior mode (sd) of $\lambda_T$ for the Student simulated datasets. }  \vspace{2mm}
\centering
\footnotesize
\ra{1.3}
\tabcolsep=0.06cm
\begin{tabular}{@{}llcccccccccc@{}}\toprule
\phantom{a} & t$_{2}(ncp=-1)$  && \phantom{a} &Prior\tnote{1}& \phantom{ab} & Modulus & Box-Cox & YJ & Id & Dual   & Log  \\ 
\cmidrule{2-2} \cmidrule{5-5} \cmidrule{7-12}
  &\multirow{6}{*}{$n=100$} & \multirow{2}{*}{$P(T|\mathbf{y})$} && prior A && 0.93 & 0.04 & 0.01 & 0.01 & $<0.01$ & $<0.01$ \\
  && && prior B && 0.93 & 0.04 & $0.01$ & $0.01$ & $<0.01$ & $<0.01$  \\
	\multirow{12}{*}{}\\[-2ex]
	&& \multirow{2}{*}{$\log f(\mathbf{y}|T)$} && prior A && -188.89 & -192.00 & -193.11 & -193.14 & -200.69 & -240.54 \\
  && && prior B && -188.75 & -191.98 & -192.94 & -193.14 & -200.59 & -240.54 \\
		\multirow{12}{*}{}\\[-2ex]
  && \multirow{2}{*}{$\lambda_T$} && prior A &&  0.14 (0.25) & 1.48 (0.19) & 1.24 (0.10) & - & 2.05 (0.20)  & -\\
  && && prior B && 0.14 (0.25) & 1.48 (0.19) & 1.24 (0.10) & - & 2.04 (0.20) & -	\\  \hline
	
				\multirow{12}{*}{}\\[-2ex]
	  & t$_{2}(ncp=-1)$ & & \phantom{a} &Prior& \phantom{ab} & Modulus  & YJ  & Box-Cox & Dual & Id & Log  \\ 
 \cmidrule{2-2} \cmidrule{5-5}  \cmidrule{7-12}
&\multirow{6}{*}{$n=1000$} & \multirow{2}{*}{$P(T|\mathbf{y})$} && prior A && $>0.99$ & $<0.01$ & $<0.01$ & $<0.01$ & $<0.01$ & $<0.01$ \\
  && && prior B && $>0.99$ & $<0.01$ & $<0.01$ & $<0.01$ & $<0.01$ & $<0.01$ \\
			\multirow{12}{*}{}\\[-2ex]
	&& \multirow{2}{*}{$\log f(\mathbf{y}| T)$} && prior A && -2827.93 & -2938.13 & -2940.72  & -2941.46 & -3103.69 & -3460.36 \\
  && && prior B && -2827.83 & -2937.93 & -2940.83  & -2941.27 & -3103.69 & -3460.36  \\
			\multirow{12}{*}{}\\[-2ex]
  && \multirow{2}{*}{$\lambda_T$} && prior A && -0.41 (0.08)  & 1.46 (0.02) & 3.04 (0.12) &   3.04 (0.12) & - & - \\
  && && prior B && -0.41 (0.08)  & 1.46 (0.02) & 3.04 (0.12) & 3.04 (0.12) & - & -  \\
\bottomrule
\end{tabular}
\label{tbl:studentpmp}
\begin{tablenotes}
    \item[1]  prior A: Power-prior (see Section~\ref{sec:powerprior}); prior B: Unit-information normal prior (see Section~\ref{sec:observedfisher}).
  \end{tablenotes}
 \end{threeparttable}
\end{table*}

By and large, very minor differences in the marginal likelihoods  are observed under the two priors, 
indicating that prior A and B give compatible results as intended. Some  more systematic deviations may be observed in the Dual model 
where no value of the transformation parameter corresponds to the reference model of normality according to theory and especially prior B deviates considerably from normality.

\subsection{Sensitivity Analysis}

In this section, we conduct sensitivity analysis by graphically presenting 
the effect of the shape and/or rate parameters of each distribution under study on the  posterior modes $\ldt$  and \
the posterior model probabilities of each transformation family.

\begin{figure}[h!]
\centering
\includegraphics[scale=0.90]{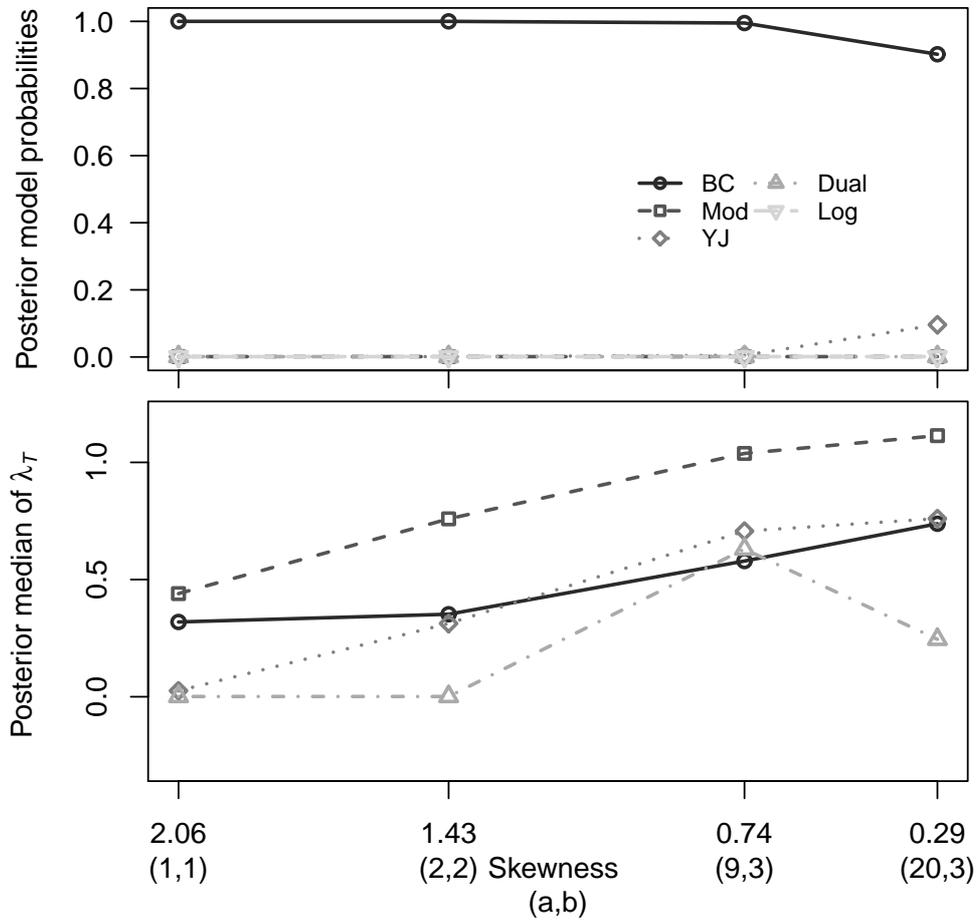}   
\caption{Posterior model probabilities $P(T|y)$ and posterior mode of  $\ldt$ under  Box-Cox, Modulus,  Yeo \& Johnson,  Dual and Log 
  according  to sample skewness for simulated Gamma(a,b) data  of $n=1000$; combinations of (a,b) are given in brackets under the skewness values.}
\label{fig:gamma_pmp}
\end{figure}

To get a more general idea as to the behavior of the best families regarding   the Gamma distribution, 
we applied the proposed methodology under prior A for various combinations of the shape and rate distribution parameters $(a,b)$ 
and for constant sample size $n=1000$. The skewness of the Gamma distribution decreases as the shape parameter increases while 
a reduction of the rate  expands the variance of the distribution. 
Figure~\ref{fig:gamma_pmp} depicts the posterior model probabilities $P(T|\mathbf{y})$ of the 
five best transformations for the Gamma distribution as a function of skewness. 
The lower part of the graph 
illustrates the posterior mode of $\ldt$ for each parametric family versus sample skewness. The associated combinations 
of the distribution parameters $(a,b)$ are also given in the horizontal axis below the skewness values. Note that 
in the first combination of $(a,b)$ values the shape parameter is taken to be unity, thus degenerating the Gamma distribution 
to an exponential distribution with  mean equal to $1/b$.
For the larger values of skewness presented, i.e. $2.0$ and $1.4$, we observe that  the Box-Cox model outperforms the rest of the transformations under consideration with posterior model probability greater than $0.99$. 
For skewness equal to $0.7$, the posterior model probabilities of the Box-Cox model  tend to  decline in contrast to the YJ model that slightly emerges for the first time with posterior model probability equal to $1\%$.
 For low skewness equal to $0.3$, the Box-Cox family is still given prominence with posterior probability $0.90$ and   the YJ family comes second
with posterior probability $10\%$. The rest of the models do not play a significant role regardless of the skewness value.
As to the posterior mode of $\ldt$ under all families except for the Dual,  we observe that it progressively increases towards unity  as the skewness decreases.
Especially for the case of  Box-Cox, 
the posterior mode of $\ldt$ is around $0.3$ for high skewness and
 increases at almost $0.7$ for very low skewness. The corresponding values of $\ldt$ for YJ 
 range from values close to zero till $0.8$.

\begin{figure}[tb]
\centering
\includegraphics[scale=0.79]{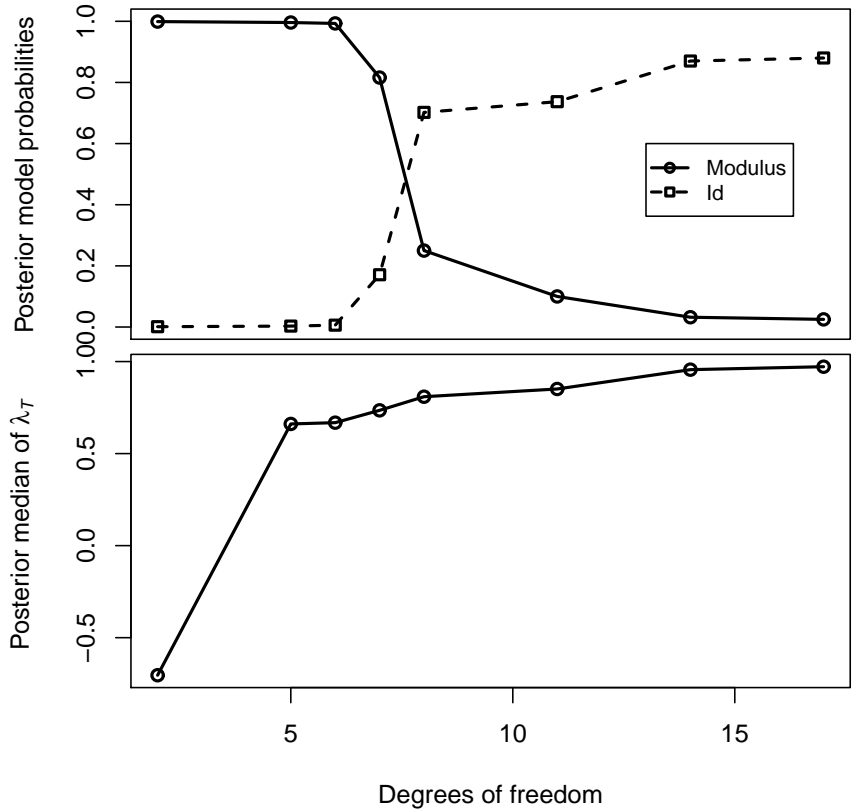}   
\caption{Posterior model probabilities $P(T|y)$  under  the Modulus and the Id transformations and posterior mode for
the Modulus $\ldt$  versus the degrees of freedom (df) for samples $n=1000$ generated from the Student distribution.}
\label{fig:student_pmp}
\end{figure}

A similar process for the Student distribution was replicated with zero non-centrality parameter.
 Figure~\ref{fig:student_pmp} provides a comparison between  the posterior probabilities of  Modulus and  Id (which are the main  competitive models in this example) versus the degrees of freedom  (df) of the distribution with constant sample size $n=1000$. For fat-tailed distributions (i.e. low degrees of freedom) the Modulus family is dominant, whereas the posterior support of Id rises as the degrees of freedom increase and the Student distribution becomes all the more similar to the normal.  The lower part of the  graph depicts the behavior of the posterior mode of  $\ldt$ under the Modulus family versus the degrees of freedom of the Student distribution, clearly showing that the posterior mode of $\ldt$ approaches unity as the degrees of freedom rise beyond a certain point. 


\section{Discussion}
\label{sec:discussion}

The  goal of this article was  to  provide a Bayesian methodology for the inference, evaluation and comparison 
of different transformation families that bring a given   dataset closest to normality. 
In our approach we consider  four parametric transformation families (Box-Cox, Modulus, Yeo \& Johnson and Dual)
 along with the standard Identity and Logarithmic transformations. The proposed methodology designates the optimal choice of transformation 
 by selecting the appropriate family and estimating the attached parameter $\lambda_T$ using Bayesian model selection.

It is made evident that the construction of reasonable priors for the  transformation families under study is fundamental 
due to the different interpretation of $\ldt$ among families. 
This issue has been dealt through the use of a power-prior approach where  
common data are generated by the reference model of the Identity transformation. 
A second prior setting, pertaining to a unit-information normal prior for $\ldt$ (or log-normal prior when it comes to Dual), was also used 
as an alternative of the first prior setting.
There was more than adequate convergence of results under both prior settings in most cases examined. 
Some differences between the two prior setups were observed only in the Dual transformation
due to the different nature and characteristics of this family.

Highly skewed data  of the Gamma distribution are sufficiently  treated  by the Box-Cox family
whereas considerable drops in the density skewness result in boosting to some extent the role of the YJ family in transforming the data.
Heavy-tailed symmetric  distributions (such as the Student and the double exponential) are  associated with the Modulus family.
In general, empirical evidence entails that the predominance of the Box-Cox transformation in the relevant literature is not always accurate 
and the selection from a wider set of transformations should become common practice.


An issue of concern for many researchers is the optimal magnitude of the shifting constant $\xi$, and more particularly of $\epsilon$, which in this article is used in the Box-Cox, the Dual and the Log transformations. 
A naive sensitivity analysis suggests generating a number of potential shifting values $\epsilon_k, k \in \mathbb{N}$ from a strictly positive Uniform distribution with large variance and check how the estimated $\ldt$ values vary according to these $\epsilon_k$ values. 
Such analysis has  been applied to a very limited scale revealing that 
the value of $\ldt$ tends to rise with the value of the shifting parameter.  
Nonetheless, a more elaborate exploration of this issue is essential. In this paper, the value of $\epsilon$ is not considered as constant and therefore as independent of the data,  but it stems from the  data itself. We also tried to derive the value of $\epsilon$ using the sample quartiles of the data as indicated in \cite{stahel_2002} but the results were discouraging in many cases.

Finally, we currently aim at extending the presented methodology  to multivariate problems.
Special interest lies on the simultaneous treatment of transformation selection along with other
aspects of modelling, such as variable selection or/and outlier detection,
using as a starting point the work presented in 
two highly motivating papers published  by 
\cite{hoeting_etal_2002}
and
\cite{gottardo_raftery_2009}.



\section*{Acknowledgements}
\label{sec:acknowledgements}

This work has received funding by the Research Committee of the National Technical University of Athens  ($\Pi$.E.B.E. 2010 Scheme).


\vspace{5 mm}


 \bibliography{biblio3}



\newpage
\vspace{10mm}
\section*{Appendix}
\appendix


\section{Calculation of the scale parameter  of prior B \\under Box-Cox}
\label{calculationofsigma}
Calculations are shown here  for the Box-Cox family. The Modulus and YJ families  follow a similar path. 
The index $T$ is kept for cohesion reasons.

Given a set of imaginary data $\ystar$ of size $\nstar$, the standard deviation ${\sigma}_{\lambda_T}$ under $T$ is based on the observed Fisher information of $\ldt$:
\be
\sigma_{\lambda_T}=
\left(-\frac{\partial^2 }{\partial \lambda_T^2} \log f(\mathbf{y}^*|\lambda_T,T)^{1/n^*} \Bigg| _{\lambda_T=1}\right)^{-\frac{1}{2}}.
\ee
Note that the observed Fisher information is evaluated at $\ldt=1$ for the Box-Cox family.
We assume without loss of generality that $\ystar$ denotes the imaginary data that have been shifted to the positive axis; in other words, instead of $(\ystar+\xi\mathbbm{1}_\nstar)$ we use $\ystar$ for simplicity reasons.
It suffices to show the calculations as to the second derivative of $\log f(\mathbf{y}^*|\lambda_T,T)^{1/n^*}$.
The  likelihood of  $\ystar$ marginalized on $\ldt$ takes the following form:
\be
f\big(\ystar|\ldt,T\big) 
\propto \big\vert J(\ystar,\ldt|T)\big\vert \cdot f\left( {\ystar}\,^{(\ldt)}|T\right).
\nonumber
\ee
Using an independent Jeffreys prior for $\left(\mu_T, \sigma^2_T \right)$, the marginal likelihood of the transformed data is:
\begin{eqnarray}
 f\left( {\ystar}^{(\ldt)}|T\right) &\propto& \left(\frac{(\nstar-1){S^*_{\mathbf{z}}}^2}{2}\right)^{-\frac{\nstar-1}{2}} \iff \nonumber\\
\log{f\left({\ystar}^{(\ldt)}|T\right)} &=& {-\frac{\nstar-1}{2}}\cdot \log{\left(\frac{(\nstar-1){S^*_{\mathbf{z}}}^2}{2}\right)} +c \nonumber,
\end{eqnarray}
with ${S^*_{\mathbf{z}}}^2$ being the sample variance of the transformed data. Under the Box-Cox family, the Jacobian is $\prod_{i=1}^\nstar{\left(y^*_i\right)^{\ldt-1}}$.
Therefore:
\begin{eqnarray}
\frac{\partial\log{f\left(\ystar|\ldt,T\right)^\frac{1}{\nstar}}}{\partial \ldt} 
&=& \frac{1}{\nstar}\sum_{i=1}^{\nstar}{\log{y^*_i}} -\frac{\nstar-1}{2\nstar} \frac{\frac{\partial\big((\nstar-1){S^*_{\mathbf{z}}}^2\big)}{\partial \ldt}}{(\nstar-1){S^*_{\mathbf{z}}}^2} \nonumber\\
&=& \frac{1}{\nstar}\sum_{i=1}^{\nstar}{\log{y^*_i}} -\frac{\nstar-1}{2\nstar} \frac{\frac{\partial\left( \sum_{i=1}^{\nstar}{\left({\left(y^*_i\right)}^{(\ldt)}-\overline{{\left(\ystar\right)}^{(\lambda_T)}}\right)^2}  \right)}{\partial \ldt}}{\sum_{i=1}^{\nstar}{\left({\left(y^*_i\right)}^{(\ldt)}-\overline{{\left(\ystar\right)}^{(\lambda_T)}}\right)^2}}.
\end{eqnarray}
If
$z_i={\left(y^*_i\right)}^{(\ldt)}=\frac{{(y^*_i)}^{\ldt}-1}{\ldt}$ and 
$\overline{\mathbf{z}}=\overline{{(\ystar)}^{(\lambda_T)}}=\frac{1}{\nstar}\sum_{i=1}^{\nstar}{z_i}$, then:
\begin{eqnarray}
\frac{\partial\left( \sum_{i=1}^{\nstar}{\left({\left(y^*_i\right)}^{(\ldt)}-\overline{{(\ystar)}^{(\lambda_T)}}\right)^2}  \right)}{\partial \ldt} 
&=&   \sum_{i=1}^{\nstar}{\frac{\partial\left[\left({\left(y^*_i\right)}^{(\ldt)}-\overline{{(\ystar)}^{(\lambda_T)}}\right)^2\right]}{\partial \ldt}} \nonumber \\
&=& \sum_{i=1}^{\nstar}{2\left(z_i-\overline{\mathbf{z}}\right) \frac{\partial\left( z_i-\overline{\mathbf{z}}  \right)}{\partial \ldt} } \nonumber \\
&=& \sum_{i=1}^{\nstar}{2\left(z_i-\overline{\mathbf{z}}\right)\left( \frac{\partial z_i}{\partial \ldt}-\frac{\partial\overline{\mathbf{z}}}{\partial \ldt}  \right).  }   \nonumber
\end{eqnarray}
In general,  $\mathbf{x}=(x_1,\ldots,x_\nstar)^T$ and the mean of the vector $\mathbf{x}$ is denoted by $\overline{\mathbf{x}}=\frac{1}{\nstar}\sum_{i=1}^{\nstar}{x_i}$.
The  derivatives of $z_i$ and $\overline{\mathbf{z}}$ with respect  to $\ldt$ are:
\begin{eqnarray}
\frac{\partial z_i}{\partial \ldt} &=&  \frac{\partial\left( \frac{ {\left(y^*_i\right)}^\ldt-1}{\ldt} \right)}{\partial \ldt} = \frac{ \frac{\partial\left[ \exp(\ldt\log{y^*_i})-1 \right]}{\partial \ldt}\ldt-\left[{\left(y^*_i\right)}^\ldt-1\right] }{\ldt^2}\nonumber\\
     &=&   \frac{ {\left(y^*_i\right)}^\ldt\frac{\partial\left( \ldt\log{y^*_i} \right)}{\partial \ldt}\ldt-{\left(y^*_i\right)}^\ldt+1 }{\ldt^2} \nonumber\\
     &=&   \frac{ {\left(y^*_i\right)}^\ldt \ldt\log{y^*_i}-{\left(y^*_i\right)}^\ldt+1 }{\ldt^2} \nonumber\\
     &=&   \frac{ {\left(y^*_i\right)}^\ldt \log{y^*_i}}{\ldt}-\frac{{\left(y^*_i\right)}^\ldt-1 }{\ldt^2} \nonumber\\
     &=&   \frac{w_i-z_i}{\ldt}  \Rightarrow \nonumber \\
\frac{\partial\overline{\mathbf{z}}}{\partial \ldt} 
     &=&  \frac{\overline{\mathbf{w}}-\overline{\mathbf{z}}}{\ldt}  \nonumber
\end{eqnarray}
where we have set $w_i={\left(y^*_i\right)}^\ldt \log{y^*_i}$. Therefore, the first  derivative  with respect to $\ldt$ is formed as follows:
\be
\dfrac{\partial\log{f\left(\ystar|\ldt,T\right)^\frac{1}{\nstar}}}{\partial \ldt} =
\dfrac{1}{\nstar}\sum_{i=1}^\nstar { \log{y^*_i} } -\frac{\nstar-1}{\nstar} \frac{ \sum_{i=1}^\nstar{(z_i-\overline{\mathbf{z}})\left( \dfrac{w_i-z_i}{\ldt}-\dfrac{\overline{\mathbf{w}}-\overline{\mathbf{z}}}{\ldt} \right)} }{\sum_{i=1}^\nstar{(z_i-\overline{\mathbf{z}})}^2}. \nonumber
\ee

We proceed with the calculation of the second  derivative of the quantity of interest with respect to $\ldt$:
\be
\dfrac{\partial^2}{\partial \ldt} 
\log{f\left(\ystar|\ldt,T\right)^\frac{1}{\nstar}}
= 0 - \dfrac{\nstar-1}{\nstar} \dfrac{\partial}{\partial \ldt}
\Bigg(  \dfrac{ \sum_{i=1}^\nstar{(z_i-\overline{\mathbf{z}})\left( \dfrac{w_i-z_i}{\ldt}-\dfrac{\overline{\mathbf{w}}-\overline{\mathbf{z}}}{\ldt} \right)} }{\sum_{i=1}^\nstar{(z_i-\overline{\mathbf{z}})}^2} \Bigg).
\label{eqn:initialder}
\ee
Moreover, we have:
\begin{eqnarray}
\sum_{i=1}^\nstar{\frac{\partial\big[(z_i-\overline{\mathbf{z}})^2\big]}{\partial \ldt}} &=&  2\sum_{i=1}^\nstar{(z_i-\overline{\mathbf{z}})}\frac{\partial(z_i-\overline{\mathbf{z}})}{\partial \ldt} =  2\sum_{i=1}^\nstar{(z_i-\overline{\mathbf{z}})}\left(\frac{\partial z_i}{\partial \ldt}-\frac{\partial\overline{\mathbf{z}}}{\partial \ldt}\right)\nonumber\\
                                           &=&  2\sum_{i=1}^\nstar{(z_i-\overline{\mathbf{z}})} \left( \frac{w_i-z_i}{\ldt}- \frac{\overline{\mathbf{w}}-\overline{\mathbf{z}}}{\ldt} \right).
																					\label{eqn:denominator}
\end{eqnarray}
Furthermore:
\begin{eqnarray}
\frac{\partial^2 z_i}{\partial \ldt^2}&=&  \frac{\partial\left(\frac{w_i-z_i}{\ldt}\right)}{\partial \ldt} = \frac{\frac{\partial(w_i-z_i)}{\partial \ldt}\ldt-(w_i-z_i)}{\ldt^2}  \nonumber\\
     &=&  \frac{\left(\frac{\partial w_i}{\partial \ldt}-\frac{\partial z_i}{\partial \ldt}\right)\ldt-(w_i-z_i)}{\ldt^2} = \frac{\left(\frac{\partial\left({y^*_i}^\ldt\log{y^*_i}\right)}{\partial \ldt}-\frac{\partial z_i}{\partial \ldt}\right)\ldt-(w_i-z_i)}{\ldt^2} \nonumber\\
     &=&  \frac{\left(\log{y^*_i}\frac{\partial\left(\exp(\ldt\log{y^*_i})\right)}{\partial \ldt}-\frac{\partial z_i}{\partial \ldt}\right)\ldt-(w_i-z_i)}{\ldt^2}  \nonumber\\
		 &=&  \frac{\left({y^*_i}^\ldt\log^2({y^*_i})-\frac{\partial z_i}{\partial \ldt}\right)\ldt-(w_i-z_i)}{\ldt^2} \nonumber \\
		 &=&  \frac{\left(w_i\log{y^*_i}-\frac{w_i-z_i}{\ldt}\right)\ldt-(w_i-z_i)}{\ldt^2} \nonumber\\
		 &=&  \frac{(\ldt\phi_i-w_i+z_i)-w_i+z_i}{\ldt^2} \nonumber \\
		 &=&  \frac{\ldt\phi_i-2w_i+2z_i}{\ldt^2}   \nonumber
\end{eqnarray}
where $\phi_i = \frac{\partial w_i}{\partial \ldt}= w_i\log{y^*_i}$. 
Therefore, given the above result, we get:
\begin{eqnarray}
\sum_{i=1}^\nstar{\left[ \left( \frac{w_i-z_i}{\ldt}- \frac{\overline{\mathbf{w}}-\overline{\mathbf{z}}}{\ldt} \right)^2 + (z_i-\overline{\mathbf{z}})\left( \frac{\phi_i}{\ldt}-\frac{2(w_i-z_i)}{\ldt^2} -\left[ \frac{\overline{\boldsymbol{\phi}}}{\ldt}- \frac{2(\overline{\mathbf{w}}-\overline{\mathbf{z}})}{\ldt^2} \right] \right) \right]}\nonumber \\
=  \sum_{i=1}^\nstar{\left[ \left( \frac{w_i-z_i}{\ldt}- \frac{\overline{\mathbf{w}}-\overline{\mathbf{z}}}{\ldt} \right)^2 + (z_i-\overline{\mathbf{z}})\left( r_i-\overline{\mathbf{r}}  \right) \right]}
\label{eqn:numerator}
\end{eqnarray}
where  $r_i= \frac{\phi_i}{\ldt}-\frac{2(w_i-z_i)}{\ldt^2}$.
Taking into consideration the results of (\ref{eqn:denominator}) and (\ref{eqn:numerator}), Equation~(\ref{eqn:initialder}) becomes:
\begin{multline*}
\frac{\partial^2\log{f\left(\ystar|\ldt,T\right)^\frac{1}{\nstar}}}{\partial \ldt^2} =\\[-2em]
\end{multline*}
\begin{align}
\hspace{3em}=& -\tfrac{\nstar-1}{\nstar} \tfrac{\sum_{i=1}^\nstar{\left[ \big(w_i-z_i-(\overline{\mathbf{w}}-\overline{\mathbf{z}})\big)^2 +\ldt^2(z_i-\overline{\mathbf{z}})(r_i-\overline{\mathbf{r}}) \right]} \sum_{i=1}^\nstar{(z_i-\overline{\mathbf{z}})^2} - 2\left[ \sum_{i=1}^\nstar{(z_i-\overline{\mathbf{z}})\big(w_i-z_i-(\overline{\mathbf{w}}-\overline{\mathbf{z}})\big)} \right]^2 } {\ldt^2\big[ \sum_{i=1}^\nstar{(z_i-\overline{\mathbf{z}})^2}  \big]^2}   
\notag
\\
=&  -\frac{\nstar-1}{\nstar} \left[ \frac{S^2_{\mathbf{w}-\mathbf{z}}+\ldt^2 S_{\mathbf{zr}}}{\ldt^2 S_\mathbf{z}^2}  -2\left(\frac{S_{\mathbf{zw}}-S_{\mathbf{z}}^2}{\ldt S^2_\mathbf{z}}\right)^2 \right].
\label{eq:finalderivative}
\end{align}

In this final expression, we have used:
\be
\sum_{i=1}^\nstar{(z_i-\overline{\mathbf{z}})(w_i-z_i-\overline{\mathbf{w}}+\overline{\mathbf{z}})} = (\nstar-1)\left(S_{\mathbf{wz}}-S^2_\mathbf{z}\right), \nonumber
\ee
\be
\sum_{i=1}^\nstar{(z_i-\overline{\mathbf{z}})\left(r_i-\overline{\mathbf{r}}\right)} = (\nstar-1)S_{\mathbf{zr}}  \nonumber
\ee
and
\be
\sum_{i=1}^\nstar{(w_i-z_i-\overline{\mathbf{w}}+\overline{\mathbf{z}})^2} = (\nstar-1)S^2_{\mathbf{w}-\mathbf{z}}  \nonumber
\ee
where		
the sample (unbiased) variance of  $\dn{\alpha}$ is denoted by $S^2_{\dn{\alpha}}$ and 
the sample covariance between  $\dn{\alpha}$ and $\dn{\beta}$ is denoted by $S_{\dn{\alpha}\dn{\beta}}$.

By substituting $\ldt=1$ in the final expression of the second derivative of $\log f(\mathbf{y}^*|\lambda_T,T)^{1/n^*}$, taking the negative of this quantity and raising it to the power of $-\frac{1}{2}$, we have the value of the scale parameter $\sigma_{\ldt}$ for the Box-Cox family.

\section{Calculation of the Scale Parameter of Prior B \\under Dual}
\label{calculationofsigmadual}

Calculations of this section pertain only to the Dual family. 
The index $T$ is kept for cohesion reasons as before.
Given a set of imaginary data $\ystar$ of size $\nstar$, the standard deviation ${\sigma}_{\log\ldt}$ under $T$ is based on the observed Fisher information of $\log\ldt$:
\be
\sigma_{\log\ldt}=
\left(-\frac{\partial^2 }{\partial (\log\ldt)^2} \log f(\mathbf{y}^*|\log\lambda_T,T)^{1/n^*} \Bigg| _{\log\lambda_T=\log\widehat{\lambda}_D}\right)^{-\frac{1}{2}}.
\ee
Note that the observed Fisher information is evaluated at $\log\widehat{\lambda}_D$ for the Dual family.
We assume as before that $\ystar$ denotes the imaginary data that have been shifted to the positive axis.

The marginal likelihood of the transformed data using an independent Jeffreys prior for $\left(\mu_T, \sigma^2_T \right)$ has been previously provided in Appendix~\ref{calculationofsigma}.
We denote the transformed parameter as $\ltil=\log\ldt$. 
Derivation is with respect to $\ltil$.
Therefore, we must redefine all relative quantities and expressions with respect to $\ltil$.
The vector of the transformed data becomes:
\be 
z_i=(y^*_i)^{\ldt}=(y^*_i)^{\exp\ltil}=
\frac{(y^*_i)^{\exp\ltil}-(y^*_i)^{-\exp\ltil}}{2\exp\ltil}.
\ee
The logarithm of the absolute Jacobian term is:
\be
\log\bigg\vert J\left(\ystar, \ltil|T\right)  \bigg\vert 
= \sum_{i=1}^{\nstar}\log\left( \frac{(y^*_i)^{\exp\ltil-1}+(y^*_i)^{-(\exp\ltil+1)}}{2} \right). 
\ee
Therefore:
\normalsize
\begin{align*}
\log{f\left(\ystar\vert\ltil,T\right)^\frac{1}{\nstar}} = &  \frac{1}{\nstar}\sum_{i=1}^{\nstar}\log\left( \frac{(y^*_i)^{\exp\ltil-1} +(y^*_i)^{-(\exp\ltil+1)}}{2} \right) \\ 
&-
\frac{\nstar-1}{2\nstar}\log\left( \frac{(\nstar-1)S^2_{\mathbf{z}}}{2}  \right),  
\end{align*}
where $\overline{\mathbf{z}}=\overline{{(\ystar)}^{(\lambda_T)}}=\frac{1}{\nstar}\sum_{i=1}^{\nstar}{z_i}$. In general,  $\mathbf{x}=(x_1,\ldots,x_\nstar)^T$ and the mean of the vector $\mathbf{x}$ is denoted by $\overline{\mathbf{x}}=\frac{1}{\nstar}\sum_{i=1}^{\nstar}{x_i}$.

We will first deal with the derivation of the Jacobian term.
The first derivative of the logarithm of the absolute Jacobian term with respect to $\ltil$ is:
\bea
\frac{\partial\log\bigg\vert J\left(\ystar, \ltil|T\right)  \bigg\vert}{\partial \ltil} 
&=&
\sum_{i=1}^{\nstar}\frac{\partial\log\left( \frac{(y^*_i)^{\exp\ltil-1}+(y^*_i)^{-(\exp\ltil+1)}}{2} \right)}{\partial\ltil} \notag\\
&=&
\sum_{i=1}^{\nstar}\frac {\frac{\partial(y^*_i)^{\exp\ltil-1}}{\partial\ltil}+\frac{\partial(y^*_i)^{-(\exp\ltil+1)}}{\partial\ltil} }
{(y^*_i)^{\exp\ltil-1}+(y^*_i)^{-(\exp\ltil+1)}}  \notag\\
&=&
\sum_{i=1}^{\nstar}\log y^*_i\exp\ltil \frac{(y^*_i)^{\exp\ltil-1}-(y^*_i)^{-(\exp\ltil+1)}}{(y^*_i)^{\exp\ltil-1}+(y^*_i)^{-(\exp\ltil+1)}}.
\eea

As to the first derivative of the quantity of interest, we have:
\begin{align}
\frac{\partial\log{f\left(\ystar|\ltil,T\right)^\frac{1}{\nstar}}}{\partial \ltil} 
= &
\frac{1}{\nstar}\frac{\partial\log\bigg\vert J\left(\ystar, \ltil|T\right)  \bigg\vert}{\partial \ltil} 
-
\frac{\nstar-1}{2\nstar} 
\frac{\frac{\partial\left( \sum_{i=1}^{\nstar}{\left( z_i-\overline{\mathbf{z}}\right)^2}  \right)}{\partial \ldt}}
{\sum_{i=1}^{\nstar}
\left( z_i -\overline{\mathbf{z}}\right)^2}
\\
= & \frac{1}{\nstar}\frac{\partial\log\bigg\vert J\left(\ystar, \ltil|T\right)  \bigg\vert}{\partial \ltil} 
- \frac{\nstar-1}{\nstar} 
 \frac{\sum_{i=1}^{\nstar}
 {\left( z_i-\overline{\mathbf{z}}\right)\left( \frac{\partial z_i}{\partial\ltil} - \frac{\partial\overline{\mathbf{z}}}{\partial\ltil}  \right)}  }
 {\sum_{i=1}^{\nstar}
 \left( z_i -\overline{\mathbf{z}}\right)^2}.
 \label{eqn:dualderiv1}
\end{align}

We are going to need the following quantities:
\bea
\frac{\partial z_i}{\partial\ltil}
&=&
\frac{\frac{\partial\left( (y^*_i)^{\exp\ltil}-(y^*_i)^{-\exp\ltil} \right)}{\partial\ltil} 2\exp\ltil 
- \left( (y^*_i)^{\exp\ltil}-(y^*_i)^{-\exp\ltil} \right) 2\exp\ltil}
{2^2\exp(2\ltil)} \notag\\
&=&
\frac{\left( (y^*_i)^{\exp\ltil}\log y^*_i \exp\ltil
+(y^*_i)^{-\exp\ltil}\log y^*_i \exp\ltil \right)
-\left( (y^*_i)^{\exp\ltil}-(y^*_i)^{-\exp\ltil} \right)}
{2\exp\ltil} \notag\\
&=&
\frac{\left( (y^*_i)^{\exp\ltil}+(y^*_i)^{-\exp\ltil} \right)\log y^*_i \exp\ltil }{2\exp\ltil}
-\frac{(y^*_i)^{\exp\ltil}-(y^*_i)^{-\exp\ltil}}{2\exp\ltil} \notag\\
&=&
\frac{\left( (y^*_i)^{\exp\ltil}+(y^*_i)^{-\exp\ltil} \right)\log y^*_i  }{2}
-\frac{(y^*_i)^{\exp\ltil}-(y^*_i)^{-\exp\ltil}}{2\exp\ltil} \notag\\
&=&
w_i-z_i
\eea

where
$w_i=\frac{\left( (y^*_i)^{\exp\ltil}+(y^*_i)^{-\exp\ltil} \right)\log y^*_i}{2}$.

\vspace{3mm}

Consequently, we have:
$\frac{\partial\overline{\mathbf{z}}}{\partial\ltil}=\overline{\mathbf{w}}-\overline{\mathbf{z}}$.
The first derivative of $w_i$ is:
\bea
\frac{\partial w_i}{\partial
\ltil}
&=&
\frac{\partial\left( (y^*_i)^{\exp\ltil}+(y^*_i)^{-\exp\ltil} \right)}{\partial\ltil}\frac{\log y^*_i}{2}
\notag\\
&=&
\left( (y^*_i)^{\exp\ltil}-(y^*_i)^{-\exp\ltil} \right)\exp\ltil\frac{\log^2 y^*_i}{2}
\notag\\
&=&
z_i\exp(2\ltil) \log^2 y^*_i \notag\\
&=&
\phi_i
.
\eea

The second derivative of $z_i$ is:
\bea
\frac{\partial^2 z_i}{\partial\ltil^2}
&=&
\frac{\partial w_i}{\partial\ltil}-\frac{\partial z_i}{\partial\ltil}
\notag\\
&=&
\phi_i-(w_i-z_i)  \notag\\
&=&
r_i
\eea
and the second derivative of the corresponding vector $\mathbf{z}$ is:
\bea
\frac{\partial^2 \overline{\mathbf{z}}}{\partial\ltil^2}
&=&
\overline{\boldsymbol{\phi}}-(\overline{\mathbf{w}}-\overline{\mathbf{z}})
\notag\\
&=&
\overline{\mathbf{r}}
.
\eea

So, (\ref{eqn:dualderiv1}) becomes:
\be
\frac{\partial\log{f\left(\ystar|\ltil,T\right)^\frac{1}{\nstar}}}{\partial \ltil} 
= 
\frac{1}{\nstar}\frac{\partial\log\bigg\vert J\left(\ystar, \ltil|T\right)  \bigg\vert}{\partial \ltil} 
-
\frac{\nstar-1}{\nstar} 
\frac{\sum_{i=1}^{\nstar}
{\left( z_i-\overline{\mathbf{z}}\right)\left( w_i - z_i - (\overline{\mathbf{w}}-\overline{\mathbf{z}})  \right)}  }
{\sum_{i=1}^{\nstar}
\left( z_i -\overline{\mathbf{z}}\right)^2}
.
\nonumber
\ee

The second derivative of the logarithm of the absolute Jacobian term with respect to $\ltil$ is given as follows:

\begin{multline*}
\frac{\partial^2\log\bigg\vert J\left(\ystar, \ltil|T\right)  \bigg\vert}{\partial \ltil^2}= \\[-1em]
\end{multline*}

\vspace{-2em}

\begin{align*}
\hspace{1cm}=&\sum_{i=1}^{\nstar}\log y^*_i
\frac{\partial \exp\ltil \frac{(y^*_i)^{\exp\ltil-1}-(y^*_i)^{-(\exp\ltil+1)}}{(y^*_i)^{\exp\ltil-1}+(y^*_i)^{-(\exp\ltil+1)}}}
{\partial\ltil}\\
=&
\sum_{i=1}^{\nstar}\log y^*_i 
\left(   
\frac{ \frac{\partial(y^*_i)^{\exp\ltil-1}\exp\ltil}{\partial\ltil} - \frac{\partial(y^*_i)^{-(\exp\ltil+1)}\exp\ltil}{\partial\ltil} }
{(y^*_i)^{\exp\ltil-1}+(y^*_i)^{-(\exp\ltil+1)}} 
\right.
\\& 
\hspace{5em} 
\left.
-\frac{\left[ (y^*_i)^{\exp\ltil-1}-(y^*_i)^{-(\exp\ltil+1)} \right]\exp\ltil 
\left( \frac{\partial(y^*_i)^{\exp\ltil-1}}{\partial\ltil} + \frac{\partial(y^*_i)^{-(\exp\ltil+1)}}{\partial\ltil}  \right)}
{\left[(y^*_i)^{\exp\ltil-1}+(y^*_i)^{-(\exp\ltil+1)}\right]^2}
\right) 
\displaybreak[1]\\
= &
\sum_{i=1}^{\nstar}\log y^*_i
\left(   
\frac{ \exp\ltil \left( (y^*_i)^{\exp\ltil-1} 
- (y^*_i)^{-(\exp\ltil+1)} \right.}
{(y^*_i)^{\exp\ltil-1}+(y^*_i)^{-(\exp\ltil+1)}} 
\right.
\\&
\hspace{5em} +\frac{ \left.  
 \left[(y^*_i)^{\exp\ltil-1}
+ (y^*_i)^{-(\exp\ltil+1)}\right]\log y^*_i \exp\ltil \right)
 }
{(y^*_i)^{\exp\ltil-1}+(y^*_i)^{-(\exp\ltil+1)}}
\\&
\hspace{5em} - \left[ (y^*_i)^{\exp\ltil-1}-(y^*_i)^{-(\exp\ltil+1)} \right]\exp\ltil \, 
\\&
\hspace{6em}   \times
\left.
\frac{ \left[ (y^*_i)^{\exp\ltil-1}\log y^*_i\exp\ltil - (y^*_i)^{-(\exp\ltil+1)}\log y^*_i\exp\ltil \right]}
{\left[(y^*_i)^{\exp\ltil-1}+(y^*_i)^{-(\exp\ltil+1)}\right]^2}
\right) 
\\
=&
\sum_{i=1}^{\nstar}\log y^*_i
\left(   
\frac
{
  \exp\ltil (y^*_i)^{2(\exp\ltil-1)}
 + \exp(2\ltil)\log y^*_i(y^*_i)^{2(\exp\ltil-1)}
 -\exp\ltil(y^*_i)^{-2}
}
{\left[(y^*_i)^{\exp\ltil-1}+(y^*_i)^{-(\exp\ltil+1)}\right]^2} 
\right.
\\&
\hspace{5em} +\frac
{
  \exp(2\ltil) \log y^*_i (y^*_i)^{-2}
 + \exp\ltil(y^*_i)^{-2} +\exp(2\ltil)\log y^*_i (y^*_i)^{-2}
}
{\left[(y^*_i)^{\exp\ltil-1}+(y^*_i)^{-(\exp\ltil+1)}\right]^2}  
\\&
\hspace{5em} +\frac
{
  -\exp\ltil (y^*_i)^{-2(\exp\ltil+1)}
 + \exp(2\ltil)\log y^*_i(y^*_i)^{-2(\exp\ltil+1)}
}
{\left[(y^*_i)^{\exp\ltil-1}+(y^*_i)^{-(\exp\ltil+1)}\right]^2}
\\&
\hspace{5em} +\frac
{
 -\exp(2\ltil)\log y^*_i(y^*_i)^{2(\exp\ltil-1)} + \exp(2\ltil) \log y^*_i (y^*_i)^{-2}
}
{\left[(y^*_i)^{\exp\ltil-1}+(y^*_i)^{-(\exp\ltil+1)}\right]^2}
\\&
\hspace{5em} 
\left.    +\frac
{
 \exp(2\ltil)\log y^*_i(y^*_i)^{-2} -\exp(2\ltil)\log y^*_i (y^*_i)^{-2(\exp\ltil+1)}
}
{\left[(y^*_i)^{\exp\ltil-1}+(y^*_i)^{-(\exp\ltil+1)}\right]^2}
\right)\\
=&
\exp\ltil\sum_{i=1}^{\nstar}\log y^*_i
\frac{\left(
(y^*_i)^{2(\exp\ltil-1)} +4\log y^*_i\exp\ltil(y^*_i)^{-2} -(y^*_i)^{-2(\exp\ltil+1)}
\right)}
{\left[(y^*_i)^{\exp\ltil-1}+(y^*_i)^{-(\exp\ltil+1)}\right]^2}
\end{align*}

The second derivative of the quantity of interest is:
\be
\frac{\partial^2\log{f\left(\ystar|\ltil,T\right)^\frac{1}{\nstar}}}{\partial \ltil^2} 
= 
\frac{1}{\nstar}\frac{\partial^2\log\bigg\vert J\left(\ystar, \ltil|T\right)  \bigg\vert}{\partial \ltil^2} 
-
\frac{\nstar-1}{\nstar} 
\frac{\partial
\frac{\sum_{i=1}^{\nstar}
{\left( z_i-\overline{\mathbf{z}}\right)\left( w_i - z_i - (\overline{\mathbf{w}}-\overline{\mathbf{z}})  \right)}  }
{\sum_{i=1}^{\nstar}
\left( z_i -\overline{\mathbf{z}}\right)^2}}
{\partial\ltil}.
\nonumber
\ee

In the  above equation, the second derivative of the absolute  Jacobian term with respect to $\ltil$ has been already calculated.  As to the second term in the above equation, by considering the relative subterms produced by applying the quotient rule of derivation, we have:
\be
N =
\sum_{i=1}^{\nstar}
{\left( z_i-\overline{\mathbf{z}}\right)\left( w_i - z_i - (\overline{\mathbf{w}}-\overline{\mathbf{z}})  \right)}
,
\ee
\be
D = 
\sum_{i=1}^{\nstar}
\left( z_i -\overline{\mathbf{z}}\right)^2
,
\ee
\bea
\frac{\partial N}{\partial\ltil} &=&  
\sum_{i=1}^{\nstar}
\left(
\frac{\partial (z_i-\overline{\mathbf{z}})}{\partial\ltil} ( w_i - z_i - (\overline{\mathbf{w}}-\overline{\mathbf{z}}))
+
(z_i-\overline{\mathbf{z}})\frac{\partial ( w_i - z_i - (\overline{\mathbf{w}}-\overline{\mathbf{z}}))}{\partial\ltil}
\right)
\notag \\
&=&  
\sum_{i=1}^{\nstar}
\left(
\left( w_i - z_i - (\overline{\mathbf{w}}-\overline{\mathbf{z}})\right)^2
+
(z_i-\overline{\mathbf{z}})\frac{\partial ( w_i - z_i - (\overline{\mathbf{w}}-\overline{\mathbf{z}}))}{\partial\ltil}
\right)
\notag \\
&=&  
\sum_{i=1}^{\nstar}
\left(
\left( w_i - z_i - (\overline{\mathbf{w}}-\overline{\mathbf{z}})\right)^2
+
(z_i-\overline{\mathbf{z}})
( \phi_i -w_i + z_i - (\overline{\boldsymbol{\phi}}-\overline{\mathbf{w}}+\overline{\mathbf{z}}))
\right)
\notag \\
&=&  
\sum_{i=1}^{\nstar}
\left(
\left( w_i - z_i - (\overline{\mathbf{w}}-\overline{\mathbf{z}})\right)^2
+
(z_i-\overline{\mathbf{z}})
( r_i  - \overline{\mathbf{r}})
\right)
,
\eea
and
\be
\frac{\partial D}{\partial\ltil}=  
2\sum_{i=1}^{\nstar}
{\left( z_i-\overline{\mathbf{z}}\right)\left( w_i - z_i - (\overline{\mathbf{w}}-\overline{\mathbf{z}})  \right)}
=2N
.
\ee
Therefore:
\begin{multline*}
\frac{\partial
\frac{\sum_{i=1}^{\nstar}
{\left( z_i-\overline{\mathbf{z}}\right)\left( w_i - z_i - (\overline{\mathbf{w}}-\overline{\mathbf{z}})  \right)}  }
{\sum_{i=1}^{\nstar}
\left( z_i -\overline{\mathbf{z}}\right)^2}}
{\partial\ltil} =   \\[-1em]  
\end{multline*}
\begin{align}
\hspace{5em}=&
\frac{
\sum_{i=1}^{\nstar}
\left(
\left( w_i - z_i - (\overline{\mathbf{w}}-\overline{\mathbf{z}})\right)^2
+
(z_i-\overline{\mathbf{z}})
( r_i  - \overline{\mathbf{r}})
\right)     \sum_{i=1}^{\nstar}( z_i -\overline{\mathbf{z}})^2
      }
{\left[
\sum_{i=1}^{\nstar}( z_i -\overline{\mathbf{z}})^2\right]^2}& 
\notag\\
&-\frac{
2\left( \sum_{i=1}^{\nstar}
{\left( z_i-\overline{\mathbf{z}}\right)\left( w_i - z_i - (\overline{\mathbf{w}}-\overline{\mathbf{z}})  \right)} \right)^2
      }
      {\left[
\sum_{i=1}^{\nstar}( z_i -\overline{\mathbf{z}})^2\right]^2}&
\notag\\
=&
\frac{S^2_{\mathbf{w}-\mathbf{z}} + S_{\mathbf{z}\mathbf{r}}}{S^2_{\mathbf{z}}}
-2(\nstar-1)^2\frac{\left( S_ {\mathbf{z}\mathbf{w}}-S^2_{\mathbf{z}} \right)^2}
{\left[ (\nstar-1)S^2_{\mathbf{z}} \right]^2}
&
\notag\\
=&
\frac{S^2_{\mathbf{w}-\mathbf{z}} + S_{\mathbf{z}\mathbf{r}}}{S^2_{\mathbf{z}}}
-2 \left( \frac{S_{\mathbf{z}\mathbf{w}}}{S^2_{\mathbf{z}}} -1 \right)^2
.
\notag\\
\end{align}
Finally, we get:
\begin{multline*}
\frac{\partial^2\log{f\left(\ystar|\ltil,T\right)^\frac{1}{\nstar}}}{\partial \ltil^2}  = \\[-3em]
\end{multline*}
\bea 
& =& 
\frac{1}{\nstar}\frac{\partial^2\log\bigg\vert J\left(\ystar, \ltil|T\right)  \bigg\vert}{\partial \ltil^2} 
-
\frac{\nstar-1}{\nstar} 
\frac{\partial
\frac{\sum_{i=1}^{\nstar}
{\left( z_i-\overline{\mathbf{z}}\right)\left( w_i - z_i - (\overline{\mathbf{w}}-\overline{\mathbf{z}})  \right)}  }
{\sum_{i=1}^{\nstar}
\left( z_i -\overline{\mathbf{z}}\right)^2}}
{\partial\ltil}
\notag\\
&=& 
\frac{1}{\nstar}
\exp\ltil\sum_{i=1}^{\nstar}\log y^*_i
\frac{\left(
(y^*_i)^{2(\exp\ltil-1)} +4\log y^*_i\exp\ltil(y^*_i)^{-2} -(y^*_i)^{-2(\exp\ltil+1)}
\right)}
{\left[(y^*_i)^{\exp\ltil-1}+(y^*_i)^{-(\exp\ltil+1)}\right]^2} 
\notag\\
&&-
\frac{\nstar-1}{\nstar}
\left( \frac{S^2_{\mathbf{w}-\mathbf{z}} + S_{\mathbf{z}\mathbf{r}}}{S^2_{\mathbf{z}}}
-2 \left( \frac{S_{\mathbf{z}\mathbf{w}}}{S^2_{\mathbf{z}}} -1 \right)^2 \right).
\eea
In this final expression, we have used:
\be
\sum_{i=1}^\nstar{(z_i-\overline{\mathbf{z}})(w_i-z_i-\overline{\mathbf{w}}+\overline{\mathbf{z}})} = (\nstar-1)\left(S_{\mathbf{wz}}-S^2_\mathbf{z}\right), \nonumber
\ee
\be
\sum_{i=1}^\nstar{(z_i-\overline{\mathbf{z}})\left(r_i-\overline{\mathbf{r}}\right)} = (\nstar-1)S_{\mathbf{zr}}  \nonumber
\ee
and
\be
\sum_{i=1}^\nstar{(w_i-z_i-\overline{\mathbf{w}}+\overline{\mathbf{z}})^2} = (\nstar-1)S^2_{\mathbf{w}-\mathbf{z}}  \nonumber
\ee
where		
the sample (unbiased) variance of  $\dn{\alpha}$ is denoted by $S^2_{\dn{\alpha}}$ and 
the sample covariance between  $\dn{\alpha}$ and $\dn{\beta}$ is denoted by $S_{\dn{\alpha}\dn{\beta}}$.

By substituting $\ltil=\log\widehat{\lambda}_D$ in the final expression of the second derivative of \\ $\log f(\mathbf{y}^*|\lambda_T,T)^{1/n^*}$, then taking the negative of this quantity and  raising it to the power of $-\frac{1}{2}$, we have the value of the scale parameter $\sigma_{\ltil}$ for the Dual family.

\end{document}